\documentclass{iopart}

\usepackage{graphicx,amssymb}

\def\la{\hbox{{\lower -2.5pt\hbox{$<$}}\hskip -8pt\raise
-2.5pt\hbox{$\sim$}}}
\def\ga{\hbox{{\lower -2.5pt\hbox{$>$}}\hskip -8pt\raise
-2.5pt\hbox{$\sim$}}}

\begin{document}

\begin{center}
\title{Implications of the cosmic ray spectrum for the mass composition at the highest energies}
\end{center}

\author{D. Allard$^{1}$, N.~G. Busca$^{1}$, G. Decerprit$^1$, A. V.\ Olinto$^{1,2}$, E. Parizot$^1$}
\address{$^1$Laboratoire Astroparticule et Cosmologie (APC), Universit\'e Paris 7/CNRS, 10 rue A. Domon et L. Duquet, 75205 Paris Cedex 13, France.}
\address{$^2$The University of Chicago, 5640 S. Ellis, Chicago, IL60637, USA}

 \date{\today}

\begin{abstract}
The significant attenuation of the cosmic-ray flux above $\sim 5\,10^{19}$~eV suggests that the observed high-energy spectrum is shaped by the so-called GZK effect. This interaction of ultra-high-energy cosmic rays (UHECRs) with the ambient radiation fields also affects their composition. We review the effect of photo-dissociation interactions on different nuclear species and analyze the phenomenology of secondary proton production as a function of energy. We show that, by itself, the UHECR spectrum does not constrain the cosmic-ray composition at their extragalactic sources. While the propagated composition (i.e., as observed at Earth) cannot contain significant amounts of intermediate mass nuclei (say between He and Si), whatever the source composition, and while it is vastly proton-dominated when protons are able to reach energies above $10^{20}$~eV at the source, we show that the propagated composition can be dominated by Fe and sub-Fe nuclei at the highest energies, either if the sources are very strongly enriched in Fe nuclei (a rather improbable situation), or if the accelerated protons have a maximum energy of a few $10^{19}$~eV at the sources. We also show that in the latter cases, the expected flux above $3\,10^{20}$~eV is very much reduced compared to the case when protons dominate in this energy range, both at the sources and at Earth.
\end{abstract}

\maketitle

\section{Introduction}

The measurement of the composition of ultra-high-energy cosmic-ray (UHECRs) and the inference of their source composition are among the main questions involved in the understanding of their enigmatic origin. The common hypothesis of a transition from a heavy composition to a light composition between $10^{17}$~eV and a few times $10^{18}$~eV is supported by the data from most cosmic ray (CR) observatories operating in this energy range \cite{Bird+93, HiRes-Mia,  HiRescomp, AugerErate}, based on the measurement of the depth of the maximum longitudinal development of the CR-induced atmospheric showers, noted $ X_{\max}$. When added to the data of the KASCADE experiment, favoring a transition to heavier and heavier composition between a few $10^{15}$~eV and $10^{17}$~eV~\cite{Kascade}, this lightening of the composition is usually interpreted as being related to the transition from a heavy galactic component above $10^{17}$~eV to an extragalactic component that is dominated by light nuclei. Current data can already be used to constrain different models of the transition from galactic to extragalactic cosmic-rays \cite{Denis2007a,  Denis2007b}, although larger statistics and a better control of the systematic uncertainties will be necessary to fully discriminate between them. The phenomenology of the transition and the interpretation of the ankle should thus be better understood in the near future.
 
Current observations do not  provide sufficient statistics on composition sensitive observables to impose strong constraints above $\sim 3\times10^{19}$ eV. On the other hand, spectral data is accumulating at the highest energies and a suppression of the high energy flux is claimed with high significance by both the HiRes and the Pierre Auger observatories~\cite{HiResGZK,AugerSpectrum}. Such a flux suppression had been predicted as a consequence of the interaction of cosmic rays with photon backgrounds and assuming either a proton or a heavier nuclei composition at the source \cite{GZK66, PSB}.

The shape of this feature and the energy at which it occurs could, in principle, constrain the composition where direct observations are still missing. In this paper we investigate the constraints placed by these observables on the source composition as well as on that expected at Earth. It is organized as follows: in section \ref{method} we discuss the relevant interaction processes between cosmic rays nuclei and photon backgrounds, in section \ref{results} we present the spectra predicted by different source compositions assumptions and compare them with the measurements from Auger and HiRes. We then study the constraints imposed by these data on the expected composition at Earth for each source model considered. Finally, in section \ref{conclusions} we summarize the results and main conclusions of this work.

\section{Modeling of nuclei interactions with photon backgrounds}

In the following sections we consider the interactions of protons and nuclei with the CMB and the infra-red, optical and ultraviolet backgrounds (hereafter IR/Opt/UV for short). Although the effect of IR/Opt/UV photons on the propagated UHECR  spectrum of pure  proton sources is almost negligible,  these additional backgrounds have a significant effect on compound nuclei propagation (and play a prominent  r\^ole in the production of cosmogenic neutrinos  \cite{DDMSS05, Denis2006}). To model the IR/Opt/UV backgrounds and their cosmological evolution, we use the latest estimate of \cite{MS05} which is based on   recent data on history of the star formation rate and the evolution of galaxy luminosity functions. Let us emphasize that the IR/Opt/UV backgrounds and their cosmological evolution (as well as the physical ingredients used to make these estimated) are at present sufficiently well constrained by experimental data and lower/upper limits on the whole energy range (see Fig.~6 and discussion in \cite{MS05}) not to represent a significant source of uncertainties in the results that follow (see also \cite{Denis2006} for a short discussion). We use IR/Opt/UV calculated at 26 different redshifts ($\Delta z=0.2$) between 0 and 5.

The interactions experienced by nuclei with photon backgrounds are different from the well known proton case. Pair production (for which we use the mass and charge scaling given in \cite{Rachen}) and adiabatic losses result in a decrease of the Lorentz factor of the UHE nucleus, whereas photodisintegration processes trigger the ejection of one or several nucleons from the parent nucleus. Different photodisintegration processes become dominant in the total interaction cross section at different energies \cite{PSB}.  The lowest energy disintegration process is the Giant Dipole Resonance (GDR) which results in  the emission of one or two nucleons and  $\alpha$ particles. The GDR process is the most relevant as it has the highest cross section with  thresholds  between 10 and 20  MeV for most  nuclei (scaling approximately as  $A^{-1/6}$, except for peculiar nuclei such as $^9Be$, deuteron or trinucleons). For nuclei with mass $A \geq 9$, we use  the theoretically calculated GDR cross sections presented in \cite{Khan04}, which take into account all the individual reaction channels (n, p, 2n, 2p, np, $\alpha$,...) and are in better agreement with data than previous treatments.  For nuclei with $A < 9$, we use the phenomenological fits to the data provided by \cite{Rachen}. Note that, as we propagate particle in the (N,Z) space, we also have to take into account $\beta$-decay in the disintegration chain. 
Around 30 MeV in the nucleus rest frame and up to the photopion production threshold, the quasi-deuteron (QD) process becomes comparable to the GDR and dominates the total cross section at higher energies.   The photopion production (or baryonic resonances (BR)) of nuclei becomes relevant  above 150 MeV in the nuclei rest frame  (e.g.,  $\sim5\times 10^{21}$ eV in the lab frame for iron nuclei interacting with the CMB), and we use  the parametrization given in \cite{Rachen} where the cross section in this energy range is proportional to the mass of the nucleus. The reference for this scaling is the deuteron photoabsorption cross section which is known in great detail.  Finally, above 1 GeV, the total cross section is dominated by the photofragmentation process which fragments nuclei into individual nucleons or low mass nuclei. For the choices of $E_{max}$ in the present work, the photofragmentation process is irrelevant and will be neglected.

\begin{figure*}[ht]\centering\hfill~\hfill\includegraphics[width=0.5\linewidth]{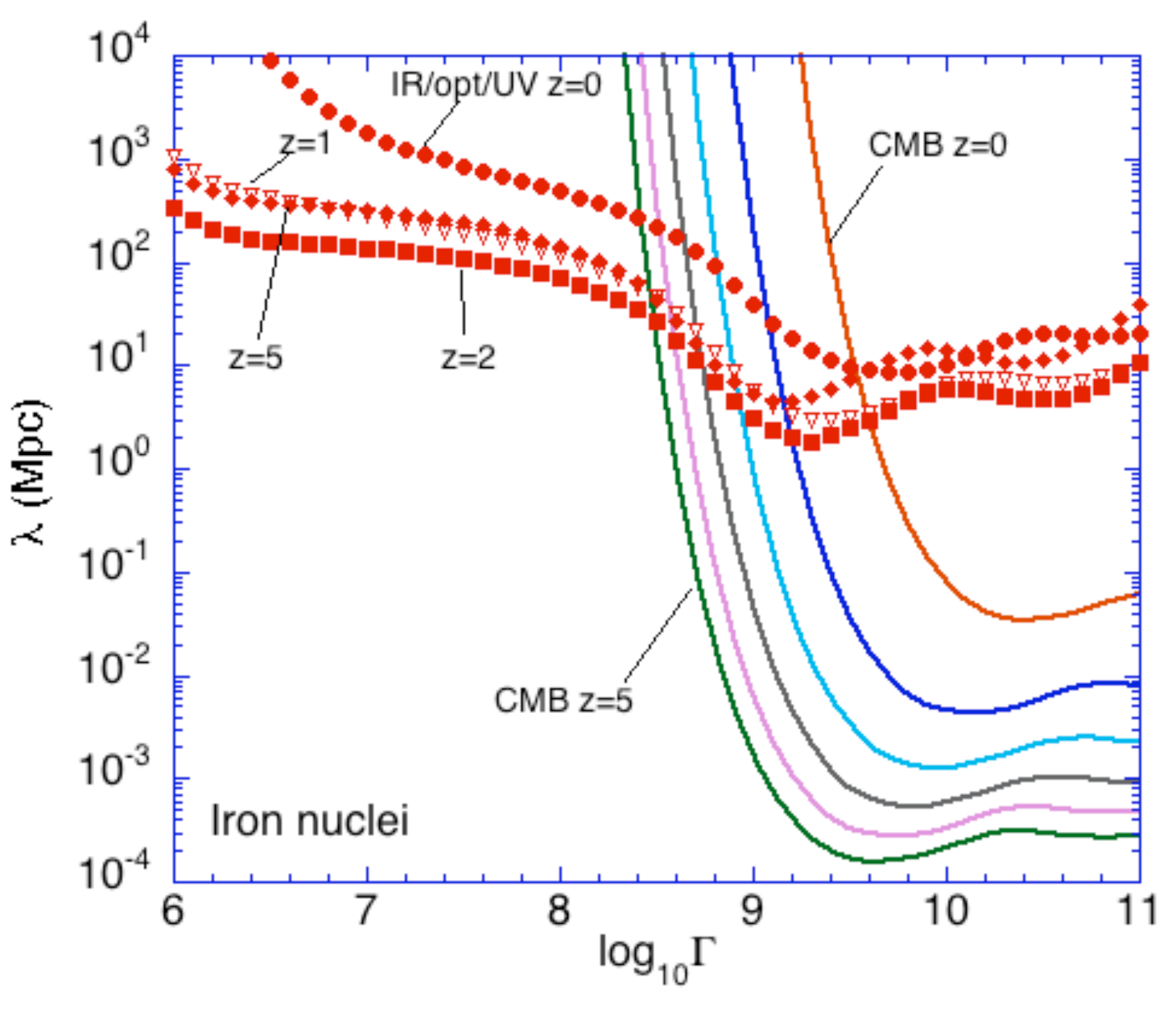}
\hfill\includegraphics[width=0.47\linewidth]{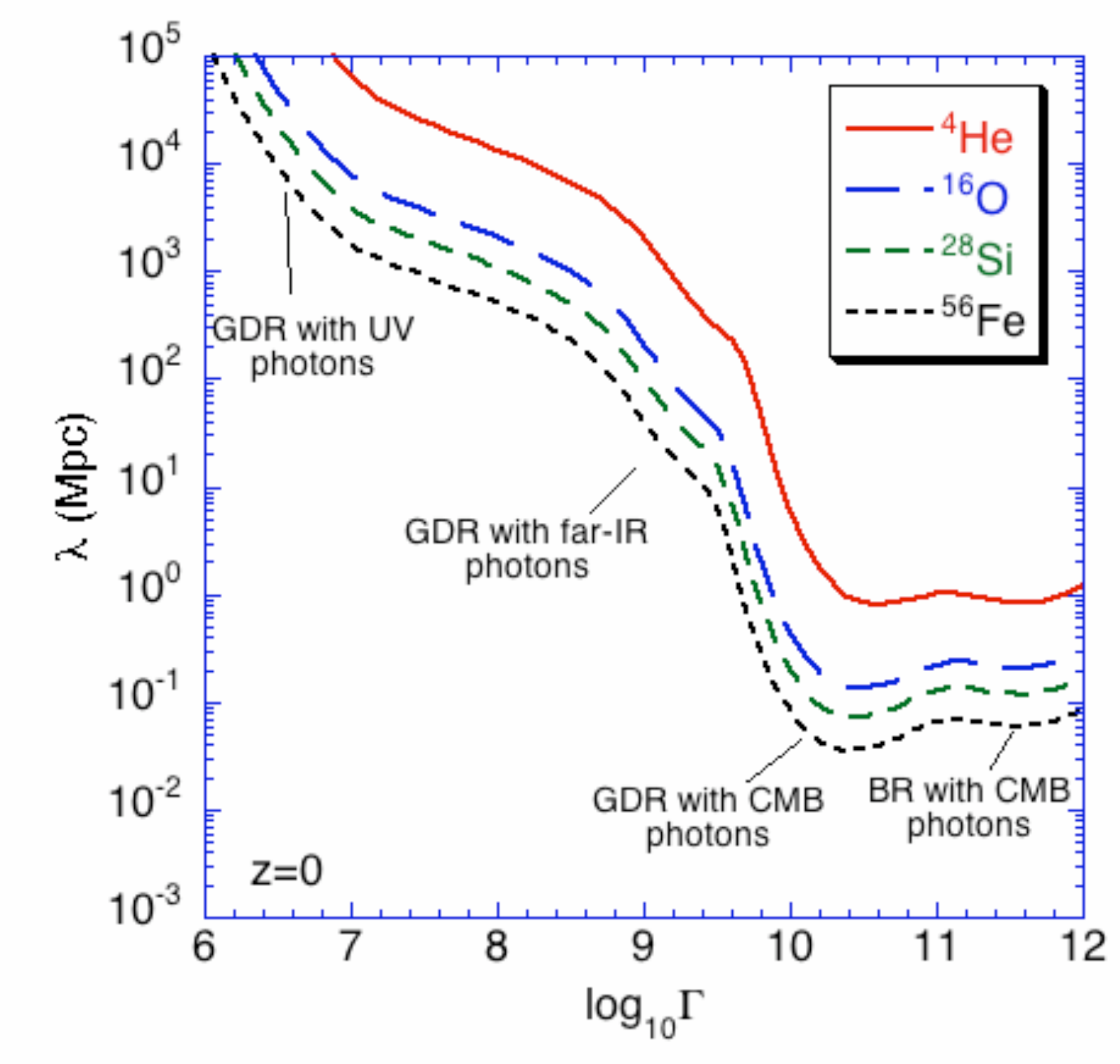}\hfill~\caption{Left: Total photodisintegration mean free path as a function of the Lorentz factor for Iron nuclei with CMB and IR/Opt/UV backgrounds at different redshift. Right: Total photodisintegration mean free path for various species (see legend) at z=0.}\label{MFP}\end{figure*}

The total photodisintegration mean free path for iron nuclei as a function of the Lorentz factor for the target CMB and IR/Opt/UV photon backgrounds is displayed in Fig.~\ref{MFP}a (see Fig.3b of \cite{Denis2006} for the contribution of the different photodisintegration processes). One can see that the mean free path evolve more slowly with redshift in IR/Opt/UV regime especially above z=1.  Fig.~\ref{MFP}b shows the comparison of the total mean free path (hereafter $\lambda$) for different nuclei. One can see several features in the energy evolution of the mean free that take place at similar Lorentz factor for all nuclei. For Lorentz factor between $\sim10^6$ and $\sim10^7$, $\lambda$ decreases as nuclei start interacting with the UV  background (via the GDR process), a slow evolution follows up to Lorentz factors $\sim10^{8.6}$ where nuclei interact with the denser far infra-red photons. $\lambda$ decreases to very low values for Lorentz factors above $\sim10^{9.5}$ where interactions with CMB photons start to contribute. The two minima in the CMB regime are due to the successive influences of the GDR and BR processes. The latter process, often omitted in nuclei propagation studies, prevents  $\lambda$ from recovering for Lorentz factors above $10^{11}$ and has a larger nucleon emission rate than the GDR process. The similarity of the threshold for the different nuclei is due to the slow scaling with the mass of the GDR threshold (the threshold difference is however visible between $^4He$ (high energy threshold) and $^{56}Fe$ in Fig.~\ref{MFP}b), whereas the scaling of $\lambda$ with A reflects the scaling of the interaction cross section (respectively approximately $\propto\,A^{7/6}$  and $\propto\,A$ for the GDR and BR processes). 

\begin{figure*}[ht]\centering\hfill~\hfill\includegraphics[width=0.32\linewidth]{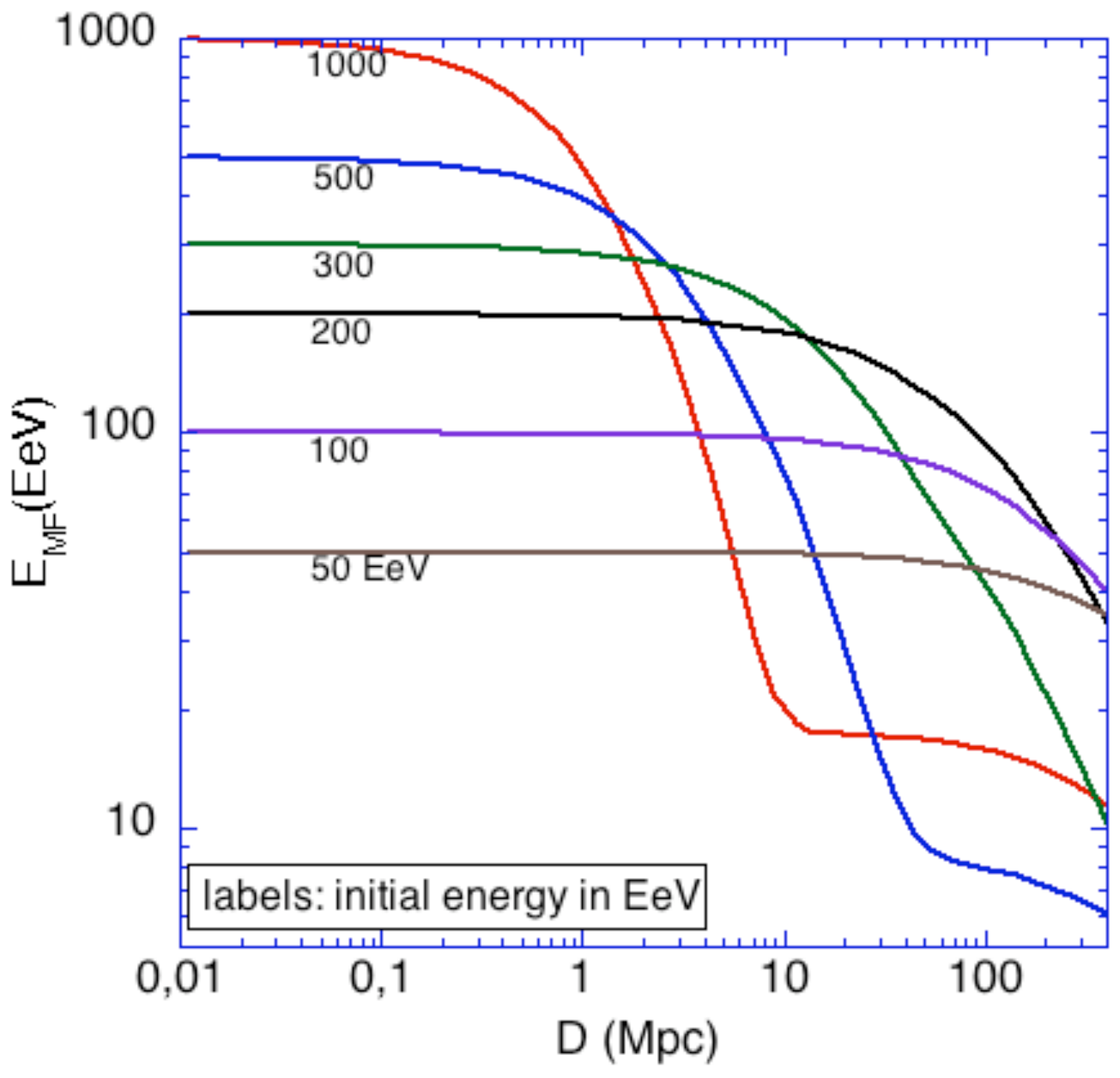}
\hfill\includegraphics[width=0.32\linewidth]{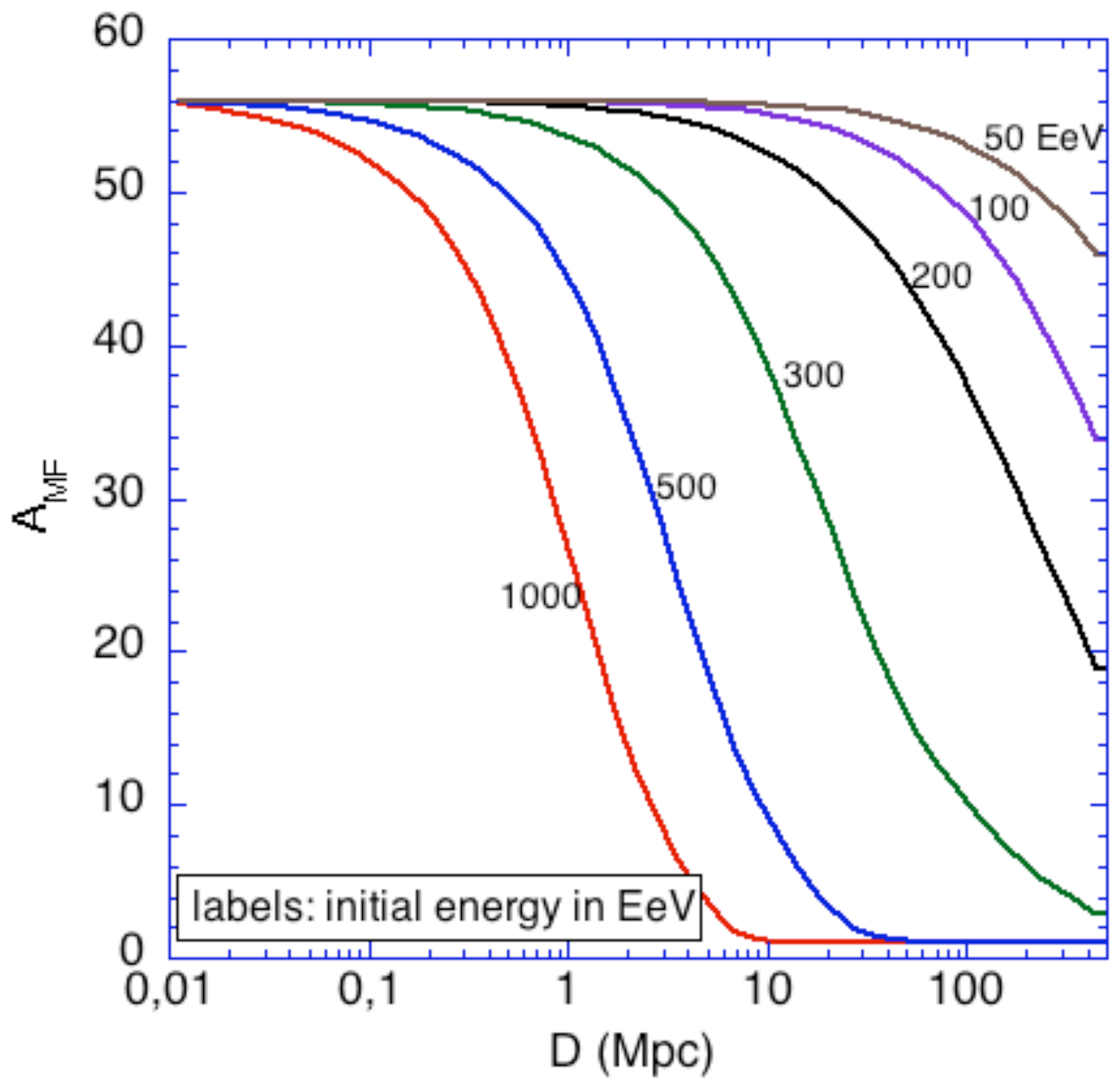}\hfill\includegraphics[width=0.32\linewidth]{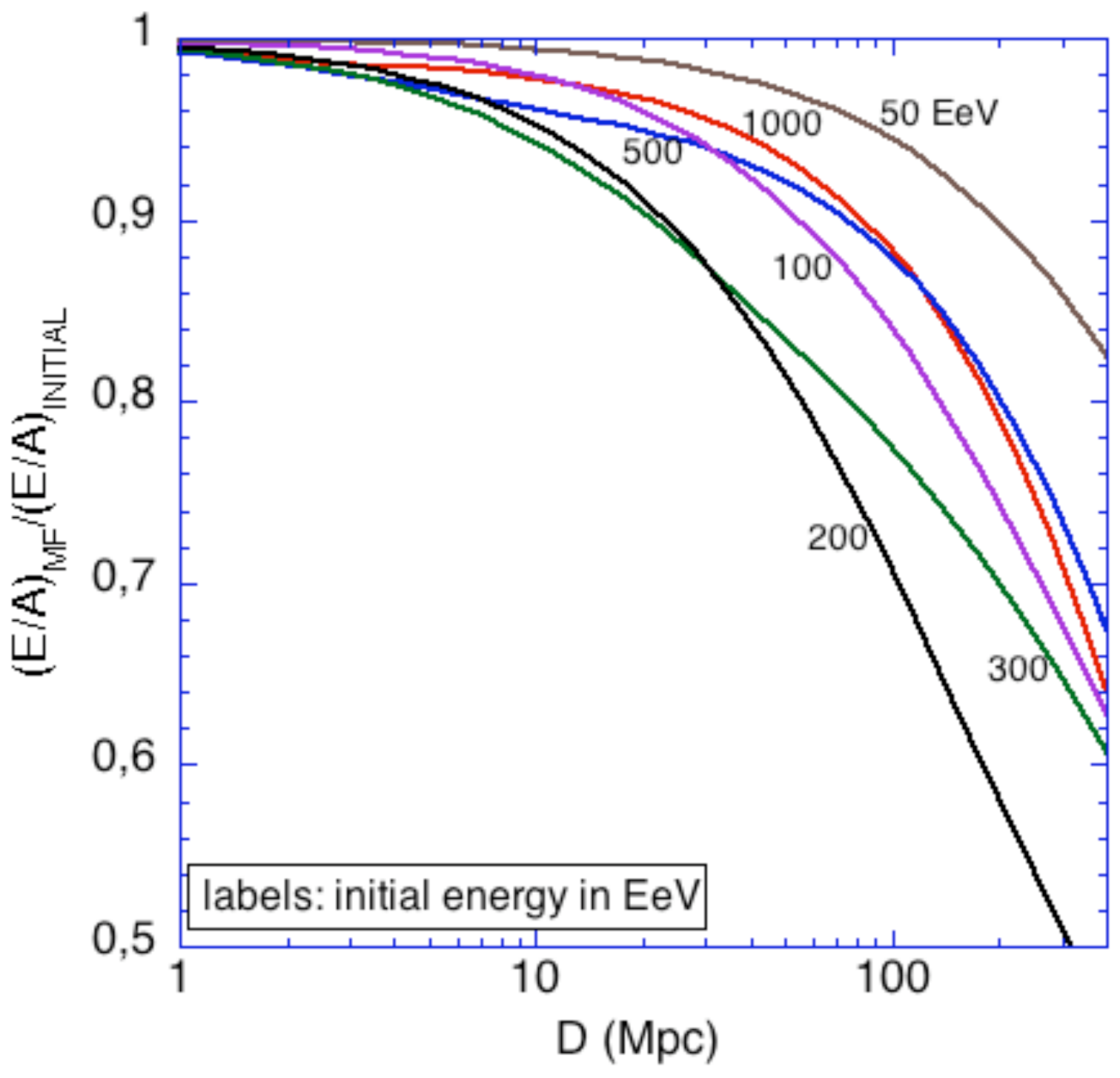}\hfill~\caption{Evolution of the energy (left), the mass (center) and the Lorentz factor (right) of the main fragment (MF) as a function of the distance traveled for iron nuclei (at the source) with different initial energies (see legend). }\label{trajectory}\end{figure*}

The photodisintegration processes as well as the pair production (with CMB and IR/Opt/UV backgrounds) and adiabatic losses are taken into account in a Monte-Carlo calculation described in more detail in \cite{Denis2006, Denis2005a}. Before turning to the predictions of spectra for various source composition assumptions, it is interesting to study the evolution of the energy, the mass and the Lorentz factor as a function of the distance travelled for different initial energies of iron nuclei at the source. This is done in Fig.~\ref{trajectory} where the mean value  of $E$, $A$ and $\gamma$ are displayed as a function of the distance traveled for initial energies of 1000, 500, 300, 200, 100 and 50 EeV with 5000 primary iron nuclei injected for each initial energy. The curves shown follow only the main nuclear fragment (i.e., the biggest fragment at the end of the photodisintegration chain). These evolutions depend on the initial redshift of the source but as we only use this exercise for a qualitative understanding of the energy losses of nuclei, we disable for the moment the cosmological evolution of the photon backgrounds in our code.

The evolution of the energy displayed on the left panel highlights  a fundamental difference between the propagation of protons and compound nuclei. One can see that nuclei with the highest initial energy lose energy faster and drop below the energy evolution of nuclei emitted with a lower initial energy (which is of course impossible in the case of protons as far the mean evolution is concerned). This is due to the fact that nuclei lose energy mainly by moving in the A-space with velocities depending on the Lorentz factor (see central panel). Mass loss mechanism that are peculiar to compound nuclei explain why the energy evolution curves can cross.


 As we pointed out above, the evolution of the mass for the different initial energies follows the evolution of the mean free path with the Lorentz factor. One can as well see that the evolution of the mass at a given initial energy becomes slower with time, when the main fragment is becoming lighter and lighter, due to the scaling of $\lambda$ with atomic mass. Moreover, the rate of nucleon losses which is extremely large for initial energies above $3\,10^{20}$ eV (CMB regime) becomes much slower when only interactions with infra-red backgorund are relevant.

The relative evolution of the Lorentz factor ($\gamma(D)/\gamma_{i}$) is less straightforward than the evolution of the energy and the mass, intuitively ordered with respect to the initial energy. The pair production mechanism indeed has a dependence in both the charge Z ($\propto\,A/Z^2$) and the Lorentz factor of the nuclei \cite{Rachen}. The rate of the decrease of the Lorentz factor is then influenced by the initial Lorentz factor and by the trajectory of the nuclei in the (A,Z) space. As a result, an iron nucleus with initial energy of 1000 EeV, that decays to proton very quickly, is less affected by pair production interactions than in the case of an initial energy around 200 EeV for which the photodisintegration process is much slower. For an initial energy around 50 EeV, the main fragment remains heavy during the full trajectory but is hardly above the threshold of pair production with the CMB. Pair production losses are then much slower than in the 200 EeV case. For the sake of completeness let us mention that the pair production interaction with CMB photons and the photodisintegration of nuclei against the far-IR bump photons occur approximately at the same energy (see Fig.~4a of \cite{Denis2006}). Interestingly, the competition between the two energy losses mechanisms is time dependent as the two photon backgrounds involved evolve differently with the redshift. This evolution is naturally taken into account in our Monte-Carlo used in the calculations that follow.

\label{method}

\section{Results}
\label{results}
\subsection{Spectrum predictions for various source composition hypotheses}
We now turn to the calculations of propagated spectra assuming various source compositions, taking into account all the relevant energy loss mechanisms described above and their time evolutions.
In what follows, we compare our predictions with HiRes \cite{HiResSpectrum} and Auger \cite{AugerSpectrum} spectra. Both experimental spectra have in common a high energy suppression of the flux but their shape are slightly different (harder in the case Auger) as well as the energy of the claimed suppressions, although both spectra are compatible within the claimed systematics. The use of both spectra will allow us to estimate wether or not the conclusions of our study are sensitive to the exact shape of the UHE spectrum and what constraints could be brought by higher statistics and lower systematics measurements in the future. To limit the number of free parameters, we will consider only source distributions with a constant comoving luminosity but will discuss the effect of a potential stronger luminosity evolution whenever needed. The source distribution is assumed to be continuous down to a minimum distance $D_{min}$ which is set to 4 Mpc unless otherwise specified. The injection spectral index $\beta$ and the maximum energy at the source $E_{max}$ (i.e., the energy above which the source spectra are exponentially attenuated) are left as free parameters to fit experimental spectra.

\begin{figure*}[ht]\centering\hfill~\hfill\includegraphics[width=0.47\linewidth]{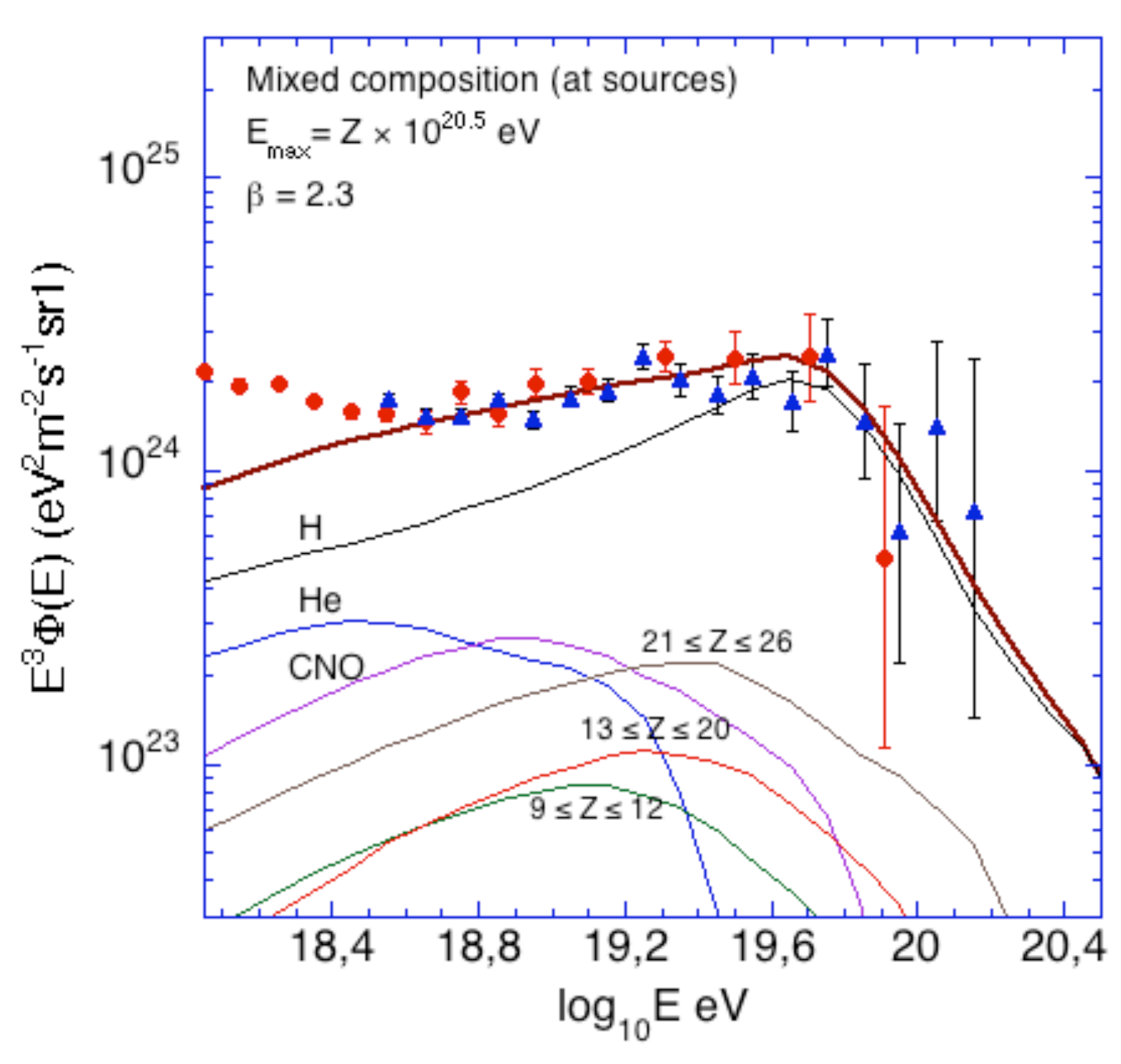}
\hfill\includegraphics[width=0.47\linewidth]{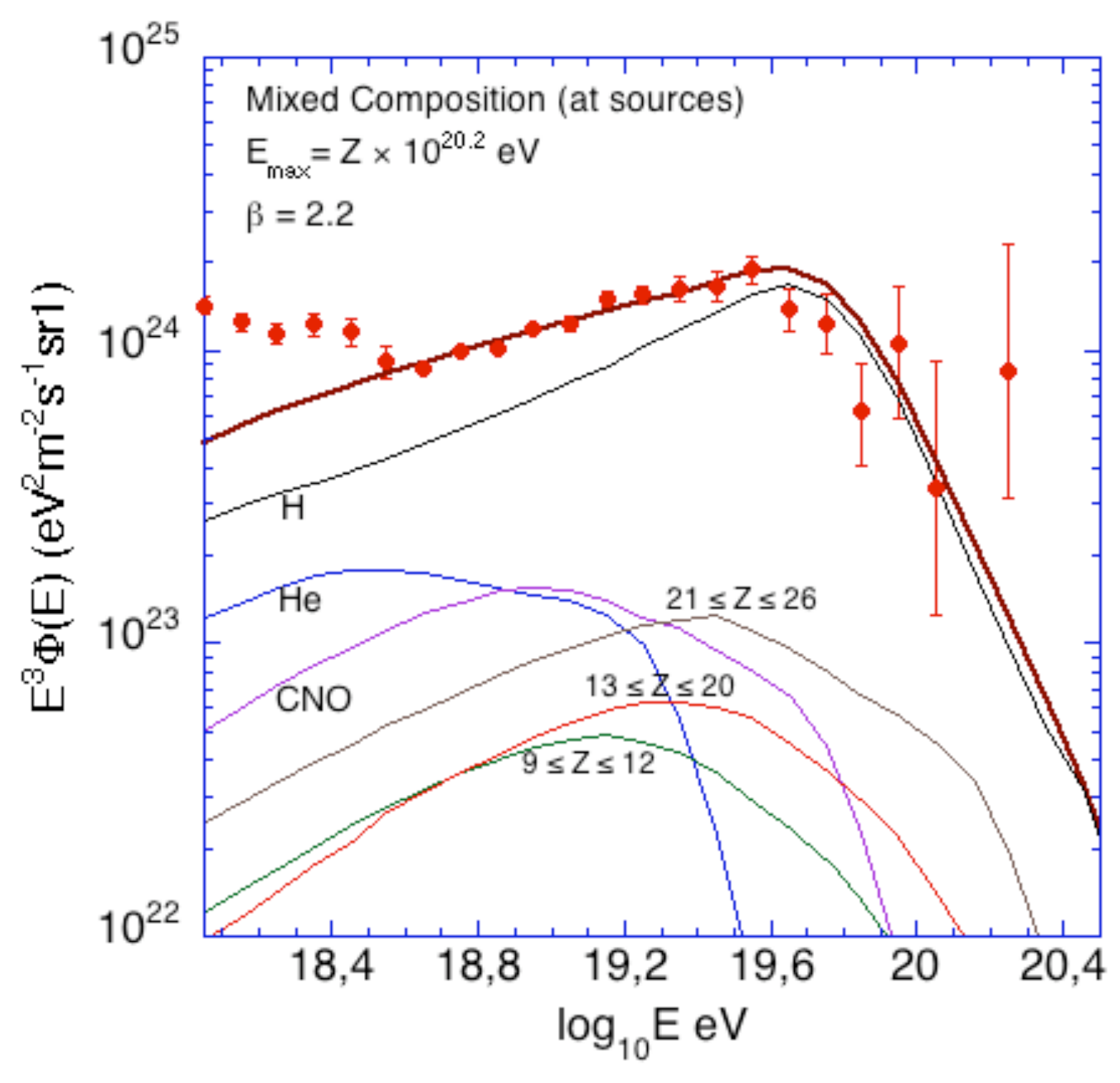}\hfill~\caption{Propagated
spectra obtained assuming a mixed source composition compared to HiRes (left) and Auger (right) spectra, the different components are displayed .}\label{MixedSpectra}\end{figure*}

We start our series of calculations by using our usual proton dominated mixed composition hypothesis (assuming the same composition as low energy galactic cosmic-rays, see \cite{Denis2007a, Denis2007b,Denis2006,Denis2005a} for more detais). The results are displayed in Fig. \ref{MixedSpectra}. Good fits can be found of both experimental spectra, with spectral index of 2.3 in the case of HiRes \cite{Denis2005a} ($\beta$=2.4 is also compatible with data \cite{Denis2008}) and 2.2 with the harder Auger spectrum (with a lower maximum energy at the source). The difference of spectral index does not have any relevant impact on the evolution of the composition or the implications on the transition from galactic to extragalactic cosmic-rays \cite{Denis2007a, Denis2007b}. Between the energy of the ankle and $\sim10^{19}$ eV the relative contribution of light (proton and He), intermediate and heavy nuclei is more or less steady. Above $10^{19}$ eV, intermediate and then heavy components drop (at energies proportional to the mass) due to interactions with far-IR photons (see Fig.~\ref{MFP}) resulting in a composition that gets lighter. At the highest energies, above $5\,10^{19}$ eV, only protons and heavy nuclei are significantly present in the composition as light and intermediate nuclei are already suppressed by the interactions with CMB photons. Due to our composition hypothesis for which the relative abundance at the sources of heavy nuclei is only $\sim10\%$, the composition is then very dominated by proton ($\sim90\%$ of the composition at the Earth) and the expected decrease of the flux at the highest energies is in all respect similar to the standard GZK feature. Note that, as we pointed out in \cite{Denis2007b, Denis2008}, between $5\,10^{19}$ and $2\,10^{20}$ eV,  the relative abundance of the heavy component increases because of the photopion interaction of protons before disappearing completely above $2-3\,10^{20}$ eV due to GDR interactions with the CMB. In this energy range the proton component actually drops faster than the heavy one (see below).

\begin{figure*}[ht]\centering\hfill~\hfill\includegraphics[width=0.47\linewidth]{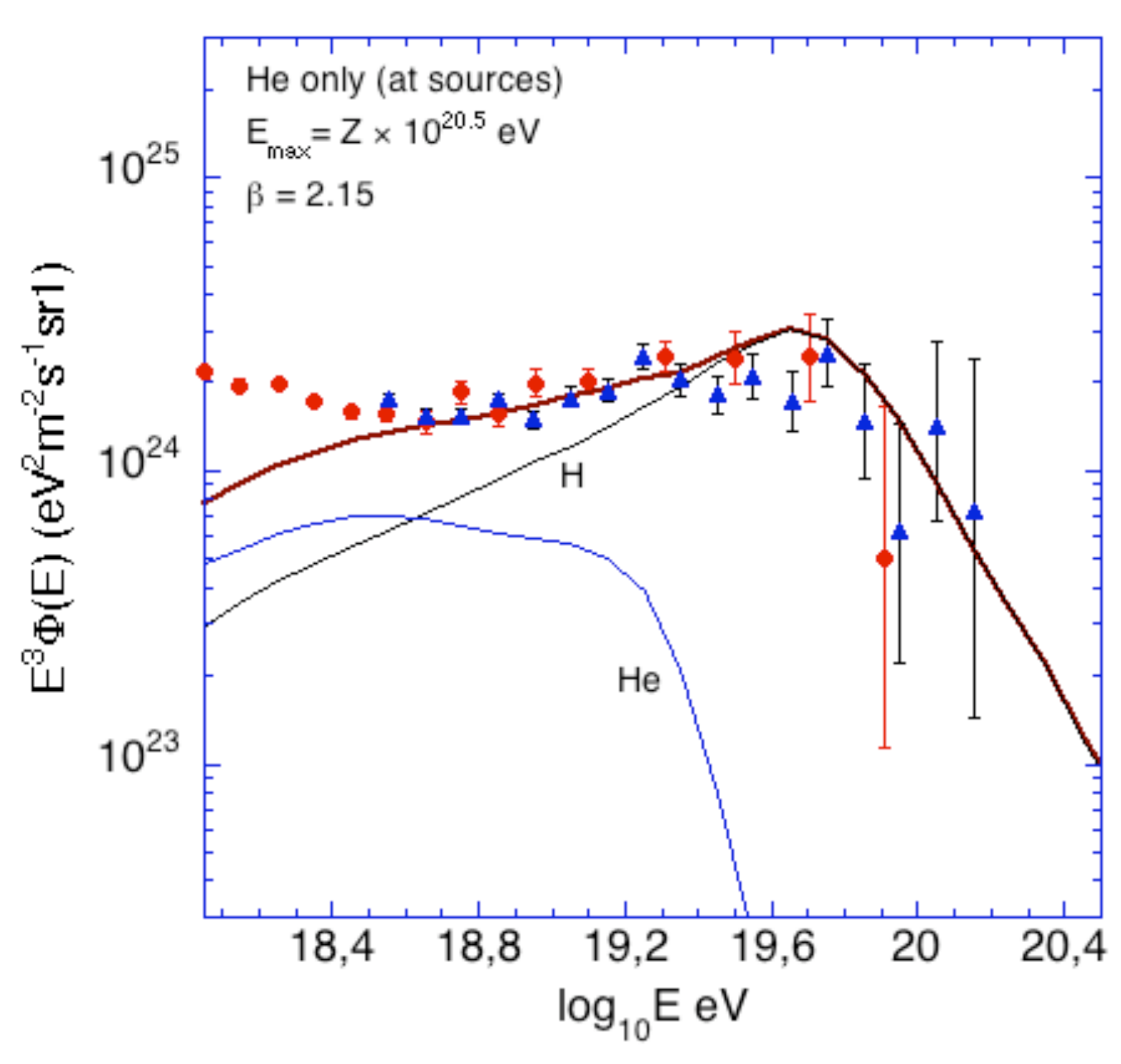}
\hfill\includegraphics[width=0.47\linewidth]{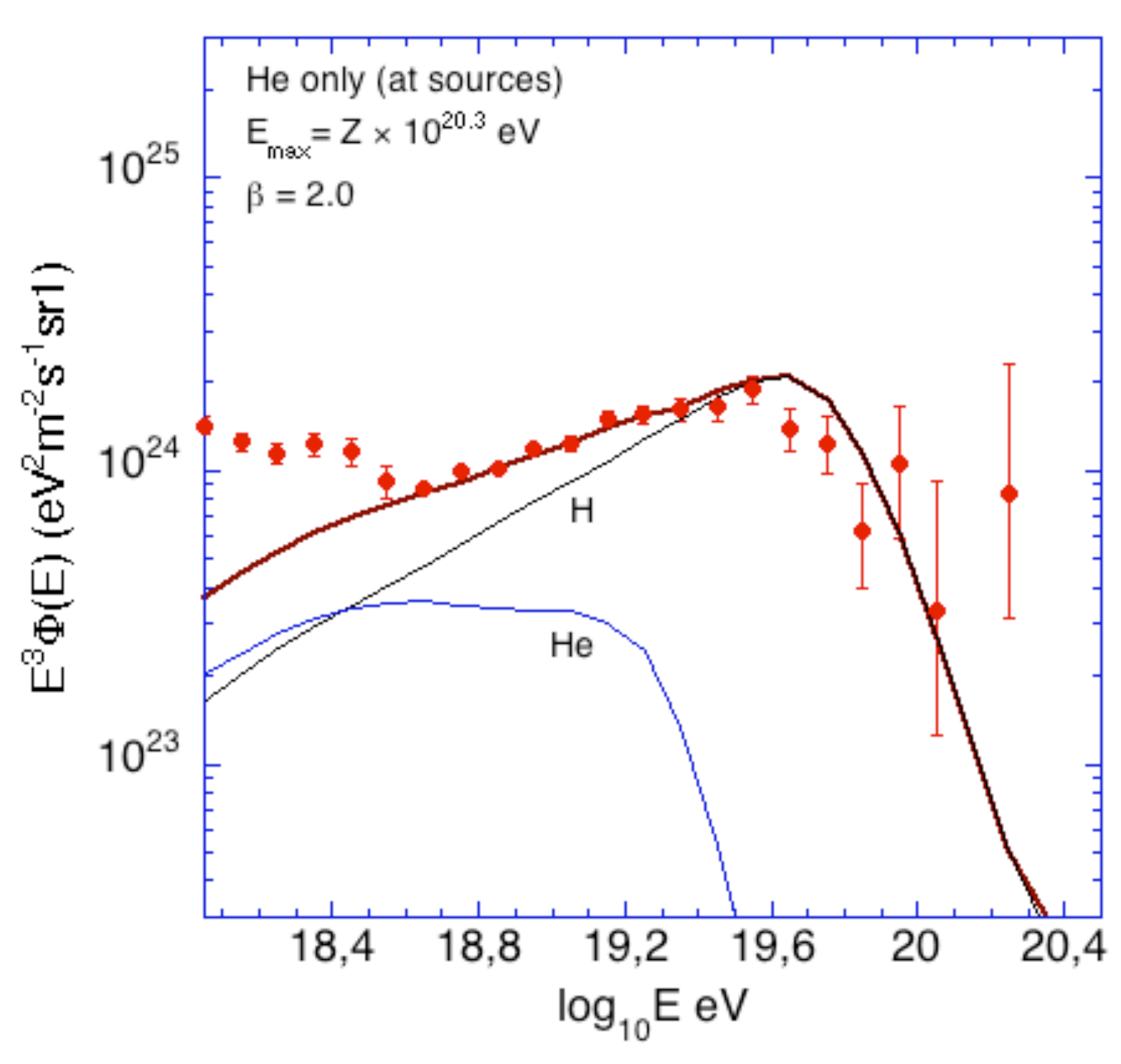}\hfill\includegraphics[width=0.47\linewidth]{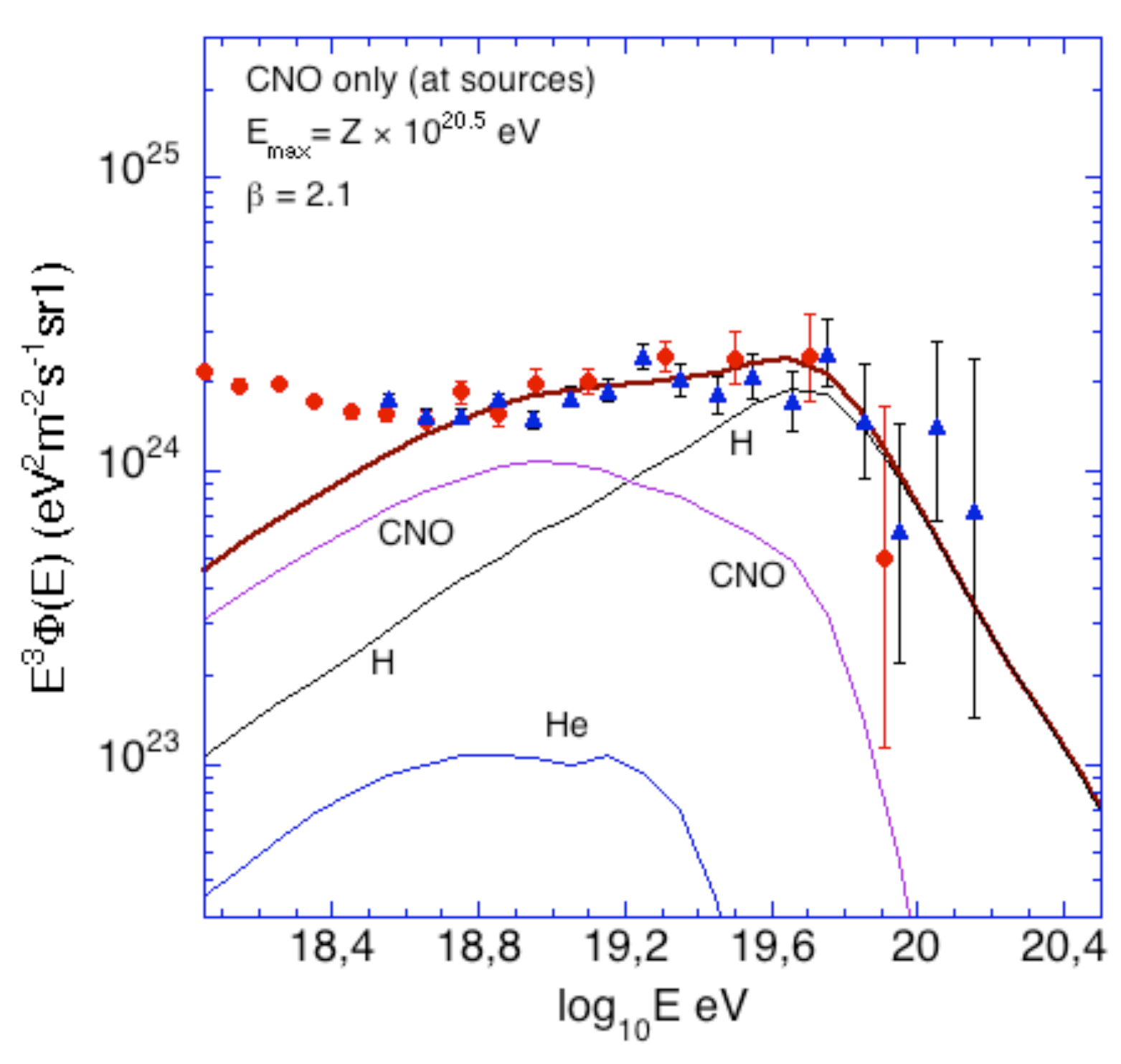}
\hfill\includegraphics[width=0.47\linewidth]{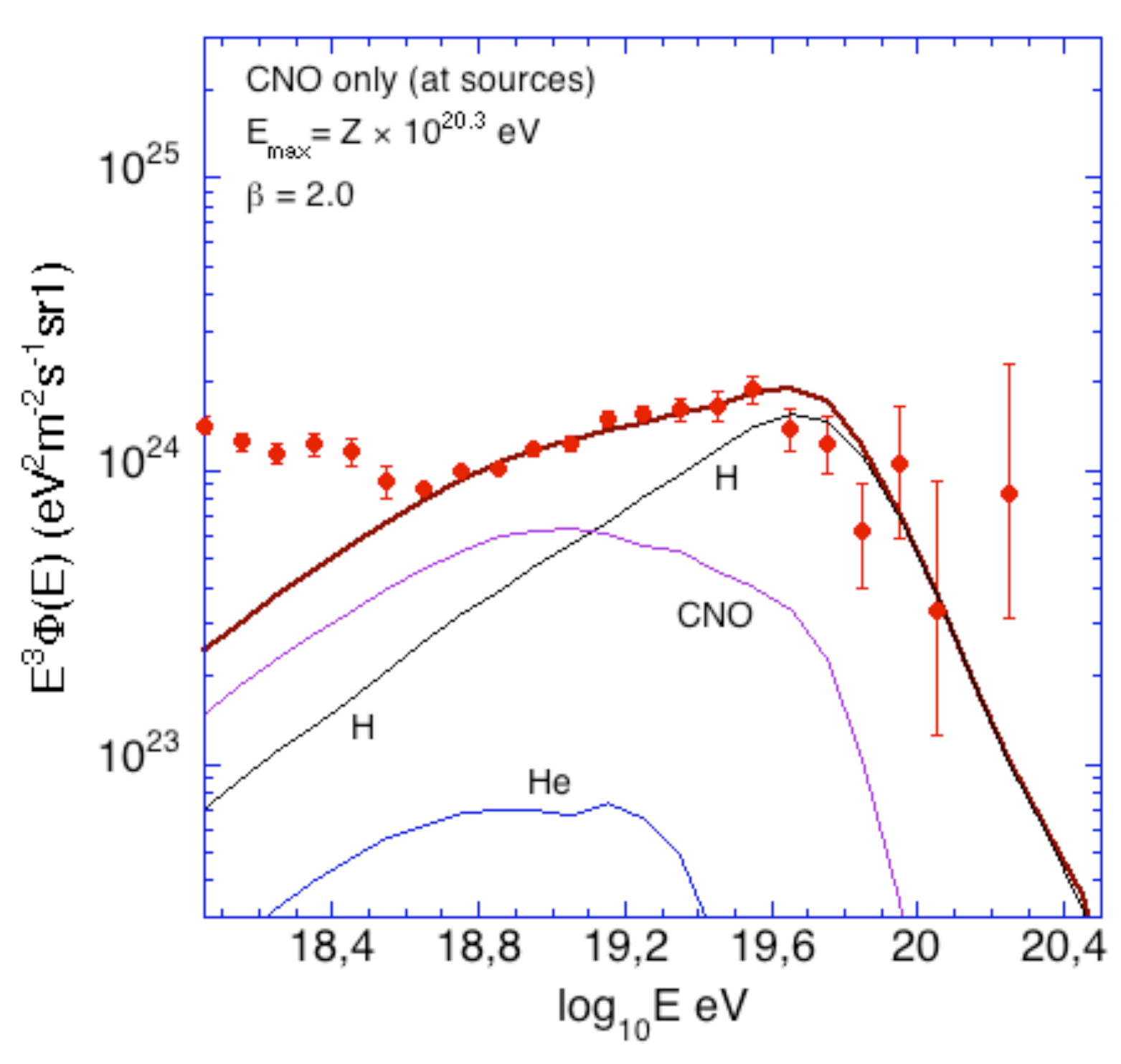}\hfill\includegraphics[width=0.47\linewidth]{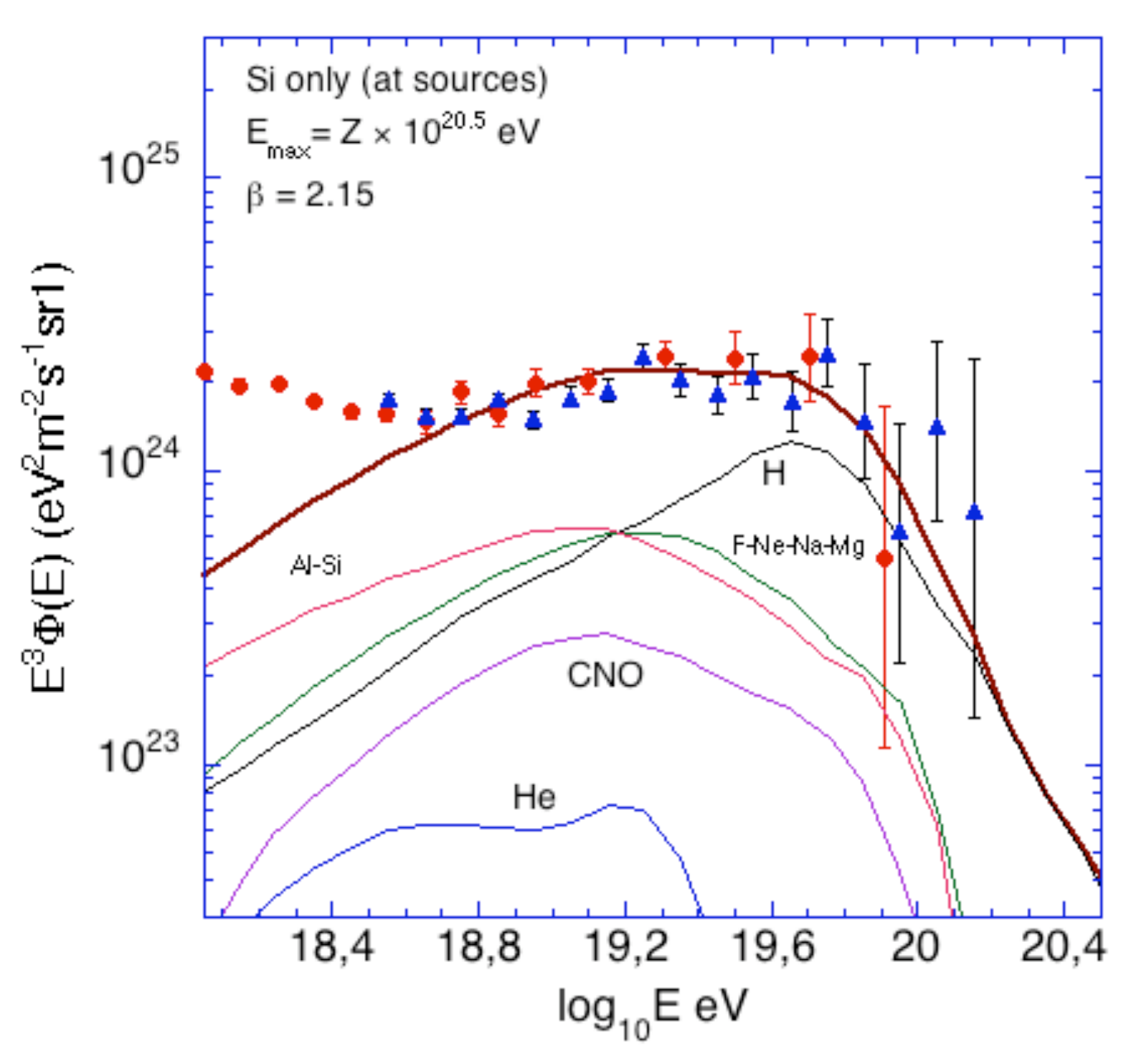}\hfill\includegraphics[width=0.47\linewidth]{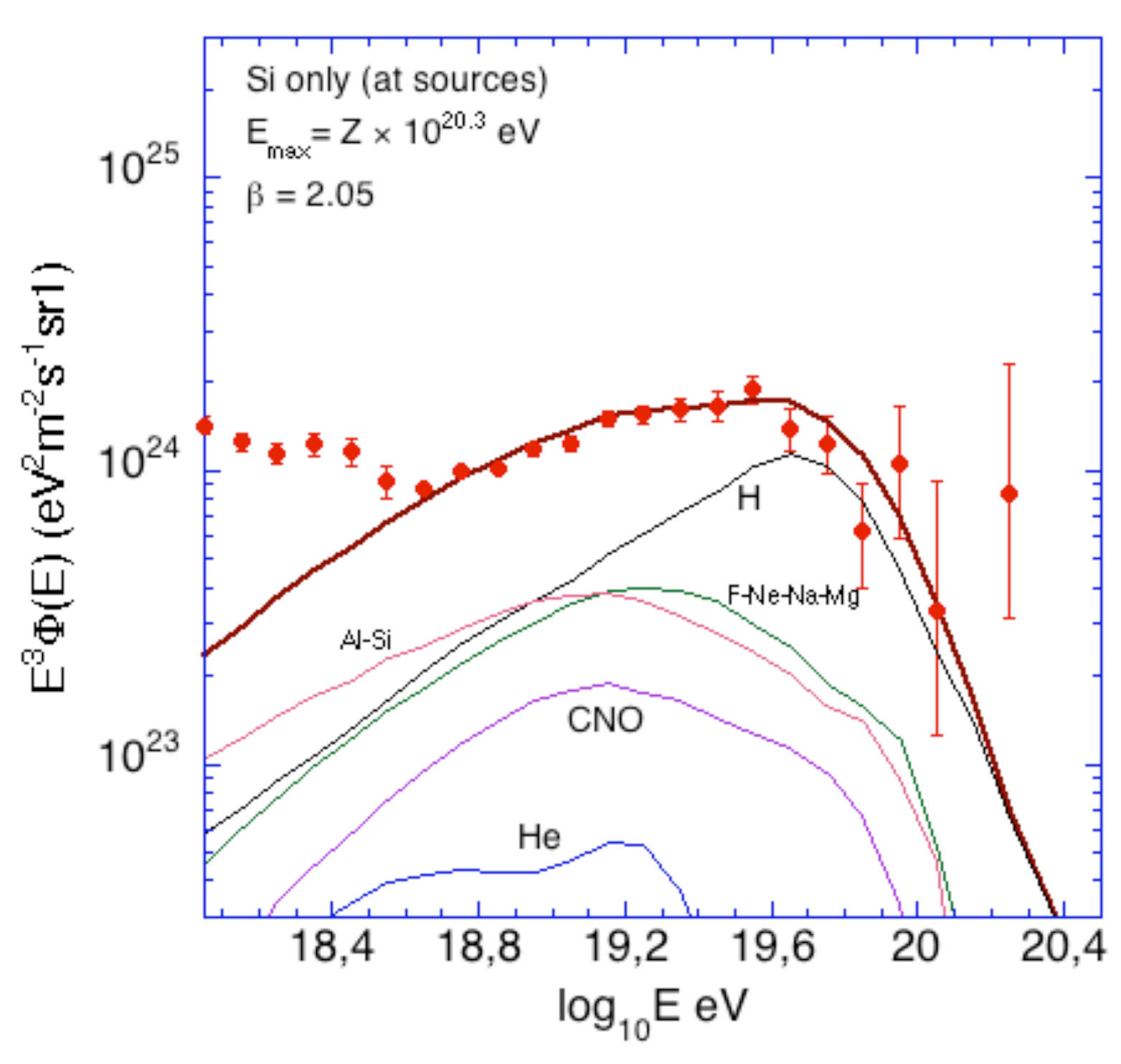}
\hfill~\caption{Propagated
spectra obtained assuming different pure source compositions, He (top), CNO (center), Si (bottom), compared with HiRes (left) and Auger (right) spectra. }\label{PureSpectra}\end{figure*}

Hydrogen is the most abundant element in the Universe as well in the galactic cosmic-ray composition, it is then natural to assume, as we did above, that it will dominate as well UHE cosmic-ray source composition. However, the sources of UHE cosmic-rays are unknown and acceleration mechanisms poorly constrained, therefore one cannot exclude the possibility that compound nuclei dominate the source composition. To study this eventuality, we propagated the injection of pure compositions of several nuclei (He, CNO and Si for light and intermediate nuclei and Fe for heavy nuclei) as in (\cite{Sigl2004, Arisaka2007, Sarkar2007}). Though totally unrealistic these hypotheses will allow us to predict the shape and composition of the spectrum at the highest energy if protons were not the most abundant nuclei at the sources.   

The propagated spectra that best fit HiRes and Auger spectra for pure He, CNO and Si source compositions are shown in Fig. \ref{PureSpectra}. As in the case of a mixed composition, the best fit spectral index are slightly harder in the case of Auger and $E_{max}$ is lower. Concerning the composition at the highest energies, in all cases a large domination of secondary protons is expected. This is due to the fact that for low and intermediate nuclei, the decrease of the primary component due to GDR interaction with far-IR photons takes place between $\sim10^{18.5}$ eV (He) and $\sim10^{19.2}$ eV (Si) above which secondary protons become dominant for the spectral indices required to fit the data. In the case of CNO and Si, the decrease of the primary component and the transition toward  the domination of secondary protons result in a mild feature in the predicted spectrum that however can not be tested at the present level of statistics. At the highest energies, the shape of the spectrum is very difficult to distinguish from the pure proton in the He and CNO cases, whereas it appears to be slightly flatter in the case of Si due to the fact that the primary component extends its influence to higher energy. In all the cases however the predictions are compatible with data. In the cases of light and intermediate nuclei dominated composition, the domination of the secondary proton component is then needed to be compatible with observations. Let us note that in the case of pure He composition we end up with a composition of 100$\%$ protons above $\sim4\,10^{19}$ eV. This extreme example shows that for light or intermediate source composition, the spectrum after propagation is expected to be rich in proton and that a large proton abundance observed above a few $10^{19}$ eV would not strongly constrain the source composition as far as light and intermediate nuclei are concerned.

\begin{figure*}[ht]\centering\hfill~\hfill\includegraphics[width=0.47\linewidth]{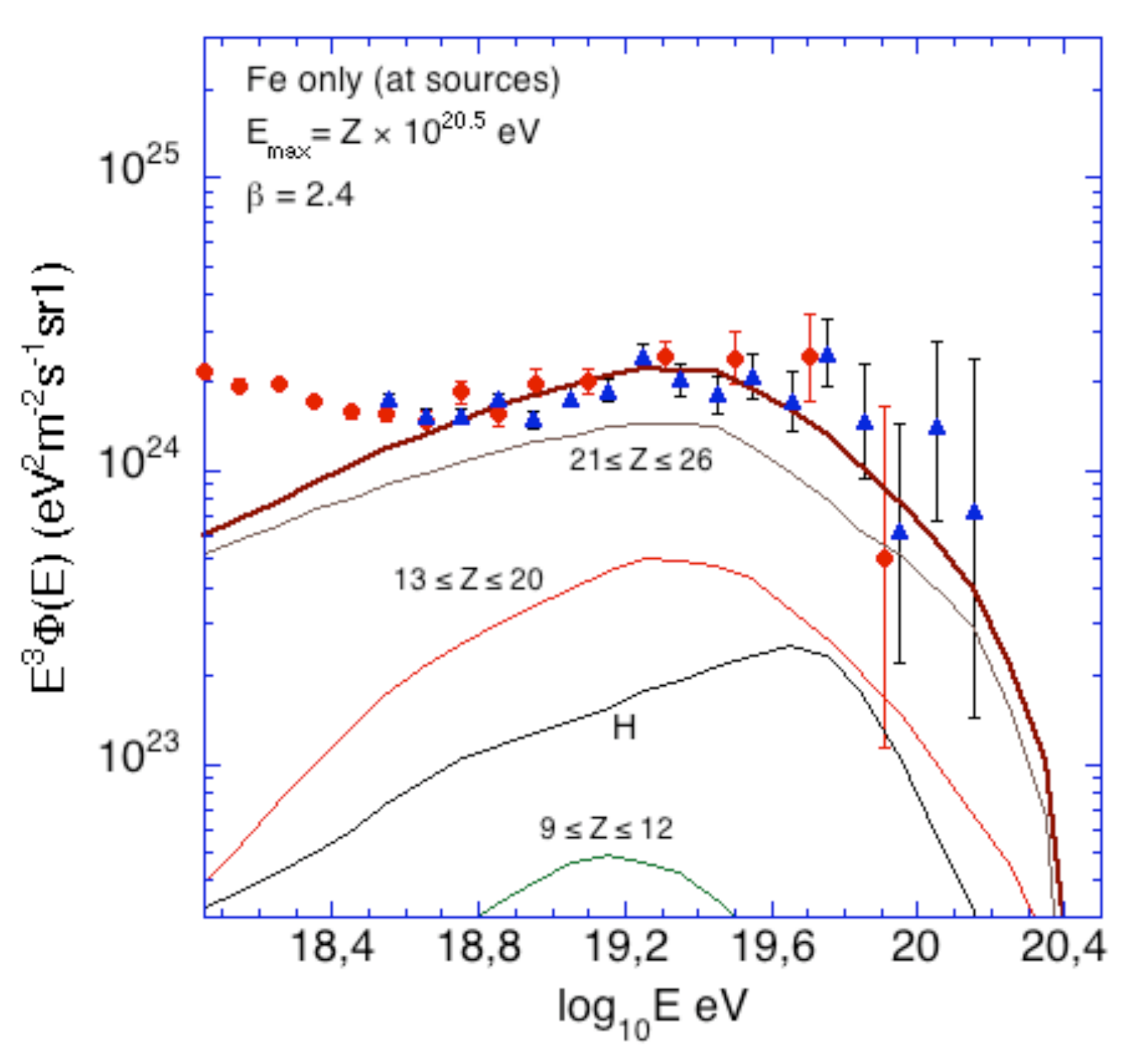}
\hfill\includegraphics[width=0.47\linewidth]{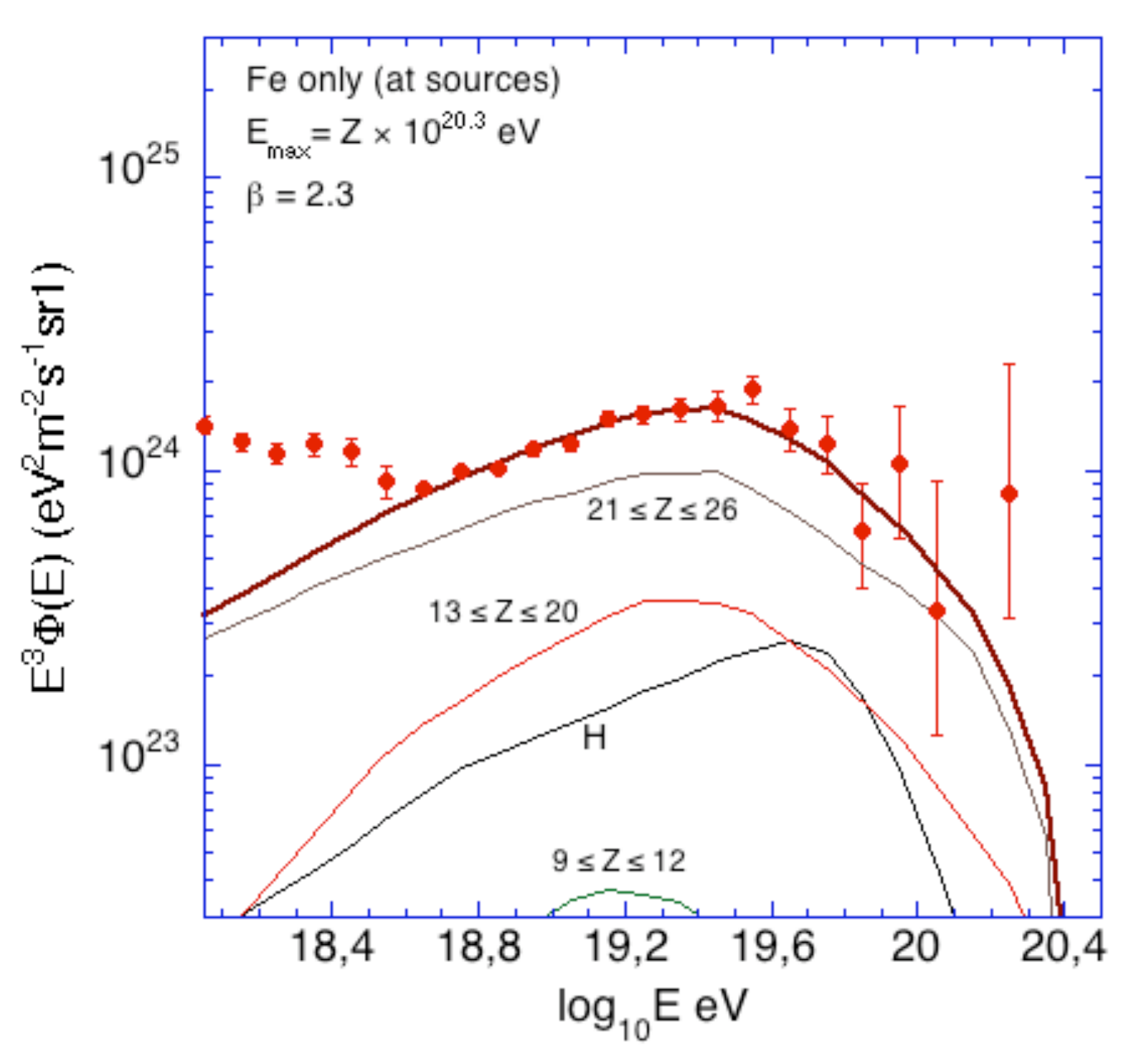}\hfill~\caption{Propagated
spectra obtained assuming a pure iron source composition compared to HiRes (left) and Auger (right) spectra.}\label{IronSpectra}\end{figure*}

The case of a heavy source composition (namely pure Fe) is displayed in Fig.~\ref{IronSpectra}. Heavy source compositions can provide good fits as well to the data ($\beta=2.4$ for HiRes and 2.3 for Auger)  though the shapes are different from the proton dominated cases. Unlike light and intermediate source compositions, one can see that the expected composition is still heavy on Earth with a quite low abundance of secondary protons. The implication of a heavy source composition on the composition at the Earth is then very different from the other cases we studied so far and the shape of the flux suppression is clearly different from the standard GZK cut-off expected for proton dominated compositions. One can see however that current data do not allow us to distinguish between the proton dominated and the heavy dominated shapes.

 Let us note that the recent claim in \cite{Berz2008} that secondary protons dominate at high energy (above $2\,10^{19}$ eV) whatever the assumed source composition  is in disagreement with our results (and to the best of our knowledge with most nuclei propagation studies, see for instance the resulting composition for pure iron sources in \cite{Sigl2004}). 
To understand these discrepancies and the implication of the source composition on the expected composition at the Earth, it is useful to study the influence of the assumed composition on the secondary proton production. 

\begin{figure*}[ht]\centering\hfill~\hfill\includegraphics[width=0.47\linewidth]{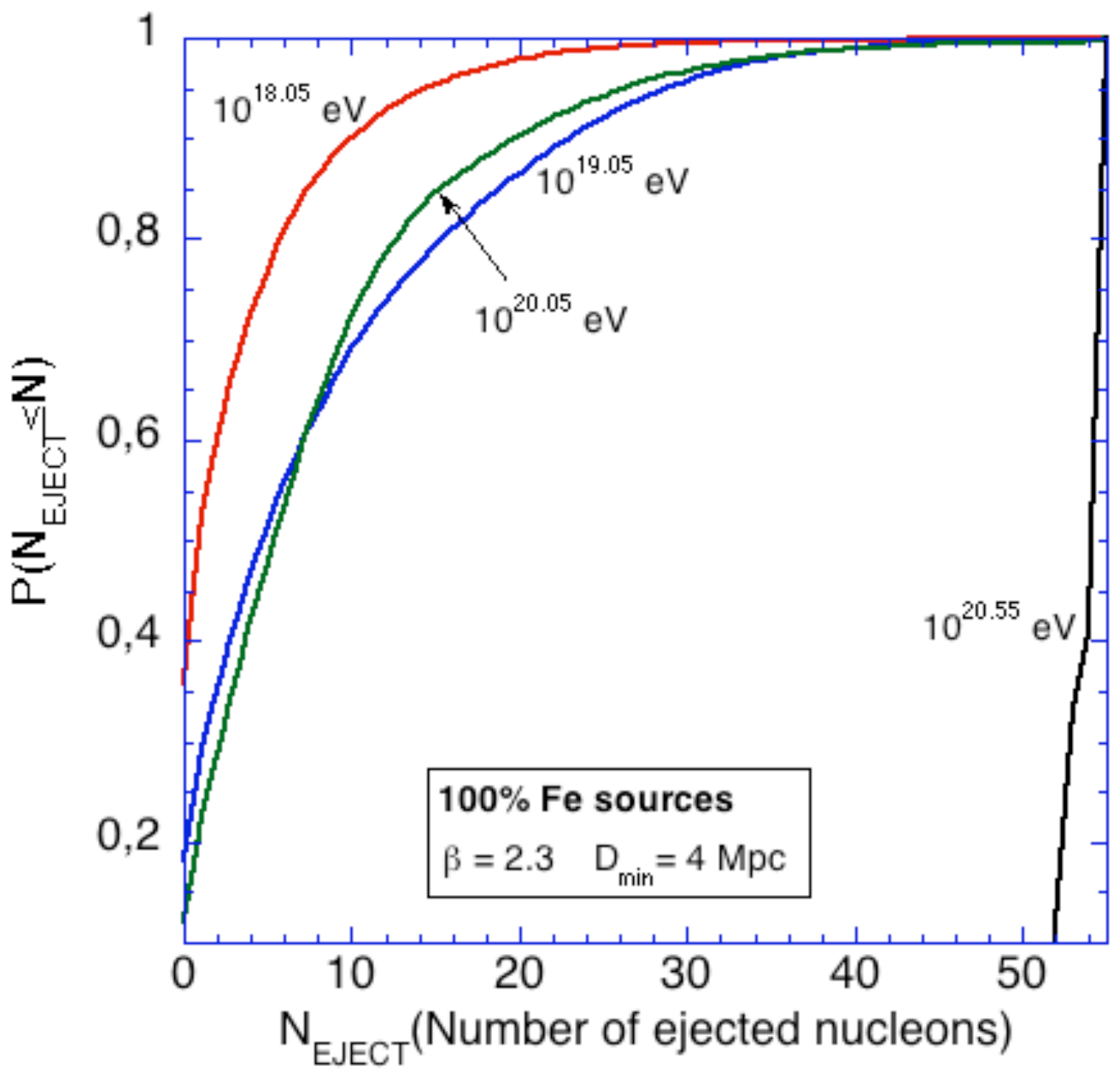}\hfill\includegraphics[width=0.47\linewidth]{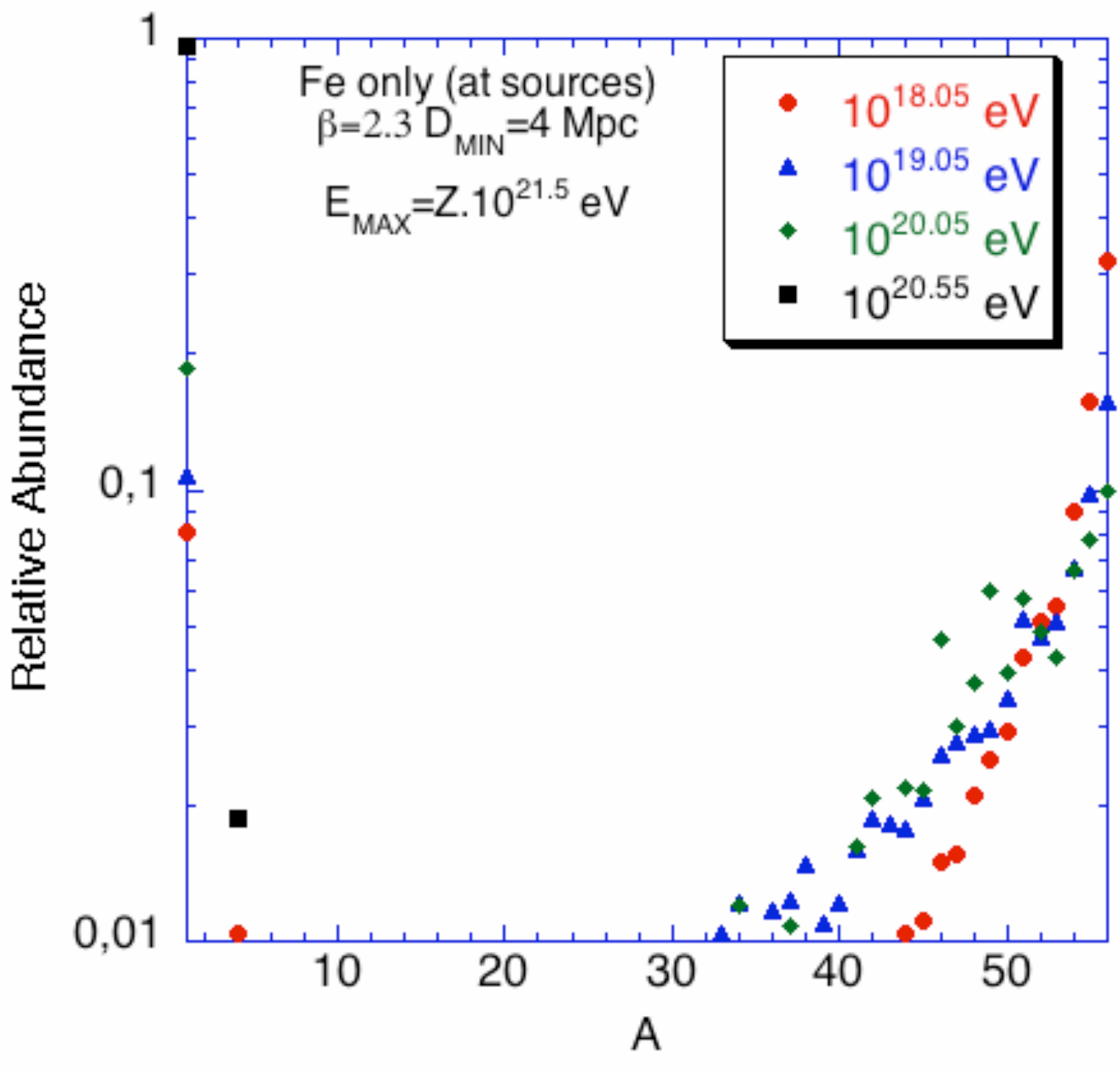}
\hfill\includegraphics[width=0.47\linewidth]{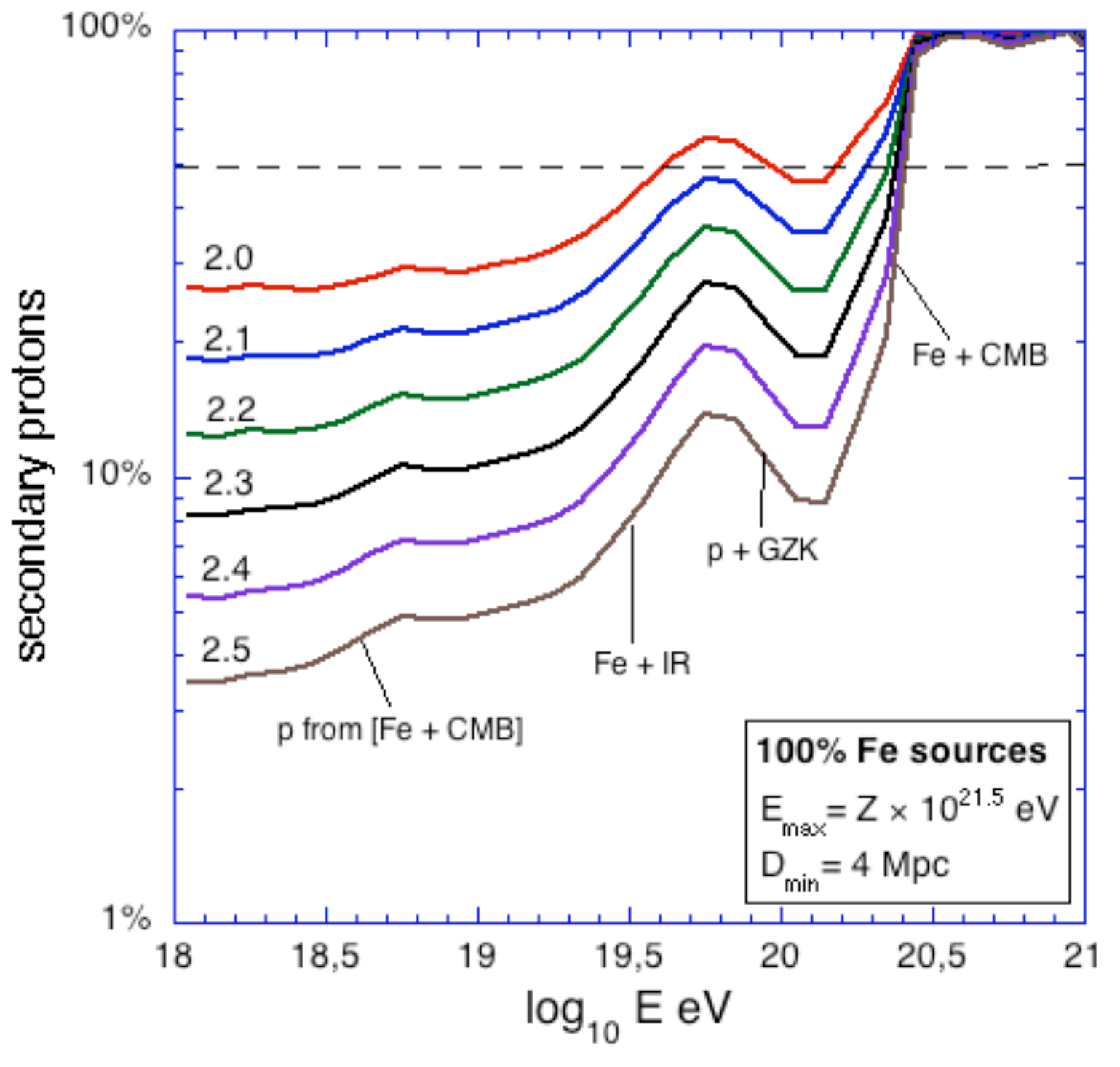}\hfill\includegraphics[width=0.47\linewidth]{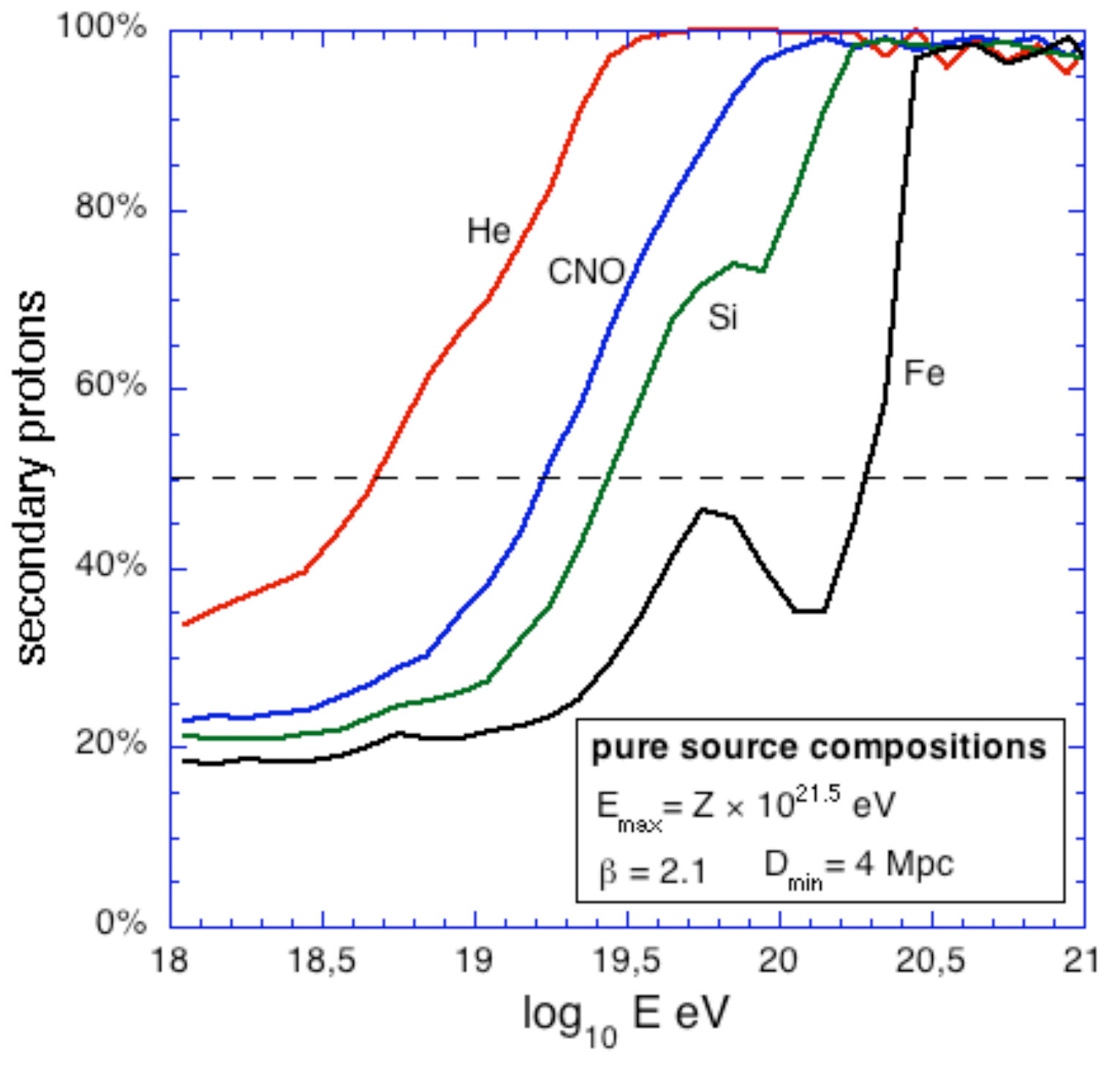}\caption{Top left: Integrated probability of the number of ejected nucleons for different energy bins (see legend). Top right: Relative abundances of the different isobar between A=1 and 56, expected on Earth in four different energy bins.
Bottom left: Evolution of the relative abundance of secondary protons at the Earth as a function of the energy in the case of a pure iron source composition for different source spectral indices. Bottom right:  Evolution of the relative abundance of secondary protons for different pure source composition hypotheses (a source spectral index of 2.1 is assumed).  }\label{Secondary}\end{figure*}

\subsection{Secondary protons production}

The different nature of the claim in \cite{Berz2008} is probably due to a very important hypothesis made by the authors in their calculation. They  assumed that once a primary nuclei interacts, it ends up fully disintegrated and results in 56 secondary protons whatever the primary energy. Though this simplifying assumption is acceptable above the energy threshold for photodisintegration with CMB photons (i.e., above $3\,10^{20}$ eV for iron), it  does not hold in the case of interactions with the lower density backgrounds (IR/Opt/UV).
To understand this point we display the integrated probability of the number of ejected nucleon (i.e., $P(N_{EJECT}\leq N$), where $N_{EJECT}$ represents the number of ejected nucleons) for the main fragment that arrives at the Earth in different energy bins. We use a 2.3 spectral index (Auger best fit) and a value of $E_{max}$ of $Z\times\,10^{21.5}$ eV (this very large value is used to maximize the number of fully disintegrated fragments arriving at the Earth for all the energy bins we consider). The result is displayed in Fig.~\ref{Secondary}a.

At low energy (see the $10^{18.05}$ or $10^{19.05}$ eV curves), the photodisintegration mean free path is very large ($\sim$1Gpc at $10^{18}$ eV and well above 100 Mpc at $10^{19}$ eV at z=0), the main fragments in these energy bins are dominated by heavy nuclei that have only lost a few nucleons or no nucleon at all as can be checked in Fig.~\ref{Secondary}b where the relative abundances of the different isobar at the same energy are displayed. The contribution of light fragments is very low as they have to be injected at a much larger energy than heavy fragment to reach the Earth at the same energy (this statement quantitatively depends on the spectral index at the sources but holds as well for harder spectral indices). When interacting with far-IR photons, the mean free path gets lower ($\sim20$ Mpc at $10^{20}$ eV) but is still sufficiently large so that heavy nuclei ($Z\geq20$) can still reach the Earth at high energy from relatively distant sources (the horizon for heavy nuclei at $10^{20}$ eV is actually slightly larger than the one of protons, see Fig.~7 of \cite{Denis2008}). Therefore heavy nuclei still dominate the distribution of main fragments (see the curve  and points for the $10^{20.05}$ eV energy bin in Fig.~\ref{Secondary}a and \ref{Secondary}b). 
In the approximation of \cite{Berz2008} all the fragments with $0<N_{EJECT}<55$ (among which a vast majority loses less than 15 nucleons, as can be seen in Fig.~\ref{Secondary}b) are wrongly considered as fully photodisintegrated and transformed into protons at lower energies. These curves clearly show that this approximation is unacceptable below the threshold of interaction with CMB photons and modify the expected balance between primary fragments and secondary photons as only nuclei that did not interact at all are kept as compound nuclei (which explains why the conclusions of \cite{Berz2008} are radically different from ours). Let us note that it can be shown as well that the full photodisintegration approximation is equally bad for all compound nuclei when interacting with photon backgrounds other than the CMB.
Finally, the full photodisintegration approximation becomes only correct above $\sim10^{20.5}$ eV when primary iron interact with CMB photons and the mean free path becomes extremely small for all the nuclei involved in the photodisintegration chain (at the minimum it goes from $\sim4\,10^{-2}$ Mpc for Fe to a bit less than 1 Mpc for He, which represent a drop of more than 2 orders of magnitude compared to the mean free path with far-IR photons (see Fig.~\ref{MFP})). Above this energy, only very light fragments arrive at the Earth provided the sources are close enough to reach the Earth before the nucleus is fully photodisintegrated. This is illustrated by the curves representing the $10^{20.55}$ eV energy bin in Fig.~\ref{Secondary}a,b. At these energies the mass of the very few non proton fragments arriving on Earth depends on the distance of the closest sources and the sample would be totally washed out of any compound nuclei if we had chosen $D_{min}\gtrsim10$ Mpc instead of 4 Mpc (see below).   

To better understand the shape of the predicted spectra for the different source composition hypotheses it is useful to study the combined evolutions of the compound nuclei and secondary proton components that can be followed by plotting the evolution of the relative abundance of secondary protons. This is illustrated in Fig.~\ref{Secondary}c where the energy evolution of the relative abundance of secondary protons in the pure iron source composition case is presented for various source spectral indices. One notices several features in the evolution that result from the features of the different components in Fig.~\ref{IronSpectra}. 

 Around $10^{18.5}$ eV, the relative abundance of protons increases as secondaries are produced more rapidly due to the interaction of primary iron with CMB photons starting around $2\,10^{20}$ eV (which triggers an increase of the number of protons produced at an energy 56 times lower). Around $10^{19}$ eV, the production of secondaries saturates as the regime of full photodisintegration is reached, while compound nuclei components at the same energy are not strongly affected by interactions, the relative abundance of secondary protons becomes more or less constant on a short energy range. The secondary protons relative abundance increases again when compound nuclei start to be affected by interactions with far-IR photons and then decreases when the secondary protons experience the GZK effect (note than the compound component decreases as well but less abruptly  than the proton component in this energy range). Finally, the compound nuclei component totally vanished between $\sim2\,10^{20}$ and $3\,10^{20}$ eV due to interactions with CMB photons. Above $3\,10^{20}$ eV, only secondaries are present provided that the maximum energy is large enough (we ensured this by choosing $E_{max}=10^{21.5}$ eV in this illustration). As mentioned before, some low mass compound nuclei can reach the Earth above this energy, depending on the distance of the closest source. Indeed, the photodisingration mean free path increases along the photodisintegration chain and the average distance needed to pass from He to  proton is $\sim$2.5 Mpc at the maximum of the interaction with CMB photons.  In any case the secondary proton abundance is expected to become very close to 1 above $3\,10^{20}$ eV. The previously described features in the evolution of the secondary protons relative abundance are present whatever spectral index we use. However, the relative abundance of secondary protons depends on the spectral index as the relative weight of the particle that can potentially be fully photodisintegrated is larger for hard spectral indices.  

When comparing, the evolution of proton secondary abundance for different source compositions ($\beta=2.1$) in Fig.~\ref{Secondary}d, one notices that the shape of the evolution and its features depend on the nature of the primary assumed. The main features seen in the case of a pure Fe composition do not apply when considering pure He composition. Indeed, He primaries produce proton secondaries at approximately  the same energy as Fe nuclei (the slight difference comes from the difference of the GDR energy threshold in the nucleus rest frame) but the energy ratio between the primary and the secondary proton is in this case only 4. As a result, He nuclei already interact with far-IR photons at the energy at which secondary protons from interaction of He nuclei with CMB photons are produced therefore the relative abundance of secondary protons is continuously increasing unlike  the case of pure Fe. The secondary protons relative abundance reaches $\sim1$ at an energy that depends on the energy threshold of the primary component  interaction with CMB photons, which is lower for He, CNO or Si than for Fe nuclei. As a result, the high energy feature related to the GZK effect on the secondary proton component does not exist for He and CNO primaries and is very faint for Si primaries.  As the energy ratio between the primary nucleus and the proton secondary is A, the relative abundance of secondary nucleons with respect to the primary component is larger for light nuclei for a given spectral index. The latter point together with the energy threshold of interaction of the primary explain the scaling of the energy of the transition between the compound dominated and secondary proton dominated phases for the different composition hypothesis as well as the shape of the different features observed in Fig.~\ref{PureSpectra}. 

\section{Discussion}
\label{conclusions}

\subsection{The highest energy end of the cosmic-ray spectrum}

In the previous section, we have calculated UHECR spectra under different hypotheses regarding the source composition and tried to derive constraints on the composition at the highest energies by comparing our predictions with experimental spectra. The conclusions that follow can be drawn from our study:

\begin{enumerate} 

\item Case of a source composition dominated by protons, or by light or intermediate mass nuclei

\begin{itemize}

\item In all the cases, i) a good fit of the data could be obtained, ii) the propagated composition, as expected to be measured at Earth, is enriched in protons at the highest energies, and iii) the expected drop in the CR flux is very similar to the GZK cut-off obtained with a pure proton composition.
 
\item In the case of a pure Si composition, however, the influence of the primary component extends to higher energies and the shape of the spectrum, although proton-dominated at the highest energies, appears different from the case of lighter nuclei sources. It nevertheless remains fully compatible with the data.

\item If heavy nuclei are present in the source composition, the proton enrichment of the propagated composition is expected to stop between the energy where the GZK effect on protons begins, namely $\sim 5\,10^{19}$~eV, and the energy where Fe nuclei start to interact  with the CMB photons, namely $\sim 2\,10^{20}$~eV. This feature is expected to leave a distinctive signature in the elongation rate at an energy that coincides with the GZK effect~\cite{Denis2007b}. If protons are accelerated above $3\,10^{20}$~eV (or produced as secondary nucleons above this energy), the composition becomes (almost) purely protons above 2--$3\,10^{20}$~eV (see \cite{Denis2007b} and below).

\end{itemize}

\item Case of heavy source compositions

\begin{itemize}

\item As illustrated in our study of the pure Fe hypothesis, the conclusions are radically different in this case. Good fits of the data can still be found, even though the shape of the high energy spectrum is different from that obtained with proton dominated compositions (but still compatible with present data), because of the influence of the photo-dissociation processes involving far-IR and then CMB photons. However, in this case the propagated composition is dominated by heavy nuclei, and the secondary proton abundance remains low.

\end{itemize}

\end{enumerate}

The behavior of heavy nuclei at the highest energies has very interesting consequences on the expected flux and composition. As mentioned earlier, above the threshold of the GDR interaction with CMB photons, the rate of nucleon losses becomes so high that only very few low mass fragments (essentially He, tri-nucleons or deuteron, since the interaction rate decreases with mass) can reach the Earth as compound nuclei above $3\,10^{20}$ eV, with the additional requirement that there is a source closer than, say, $\sim$10 Mpc. This can be seen  on Fig.~\ref{trajectory}, where it appears that an energy lower than 300~EeV is reached on average after a propagation distance of $\sim 3$~Mpc (an initial energy larger than $10^{21}$~eV would not help as the influence of the BR process is growing above this energy). As a result, in the case of a proton-dominated, mixed composition, any significant flux above $3\,10^{20}$ eV would have to be essentially purely protonic (to a good approximation). It is interesting to note that this is the only region of the spectrum where the CR flux (if any) is guaranteed to represent an almost pure proton beam. Besides, the flux above $3\,10^{20}$ eV that can be expected at the Earth crucially depends on the value of the maximum energy at the sources as well as the spatial distribution of the sources in the local universe, and it is important to note that none of these parameters are strongly constrained by current data. For the sake of completeness, let us mention that superheavy nuclei (Pb, U) would represent a potential source of  ``non protonic pollution" above $3\,10^{20}$~eV, but they would have to be vastly more abundant than in the galactic cosmic-ray composition to be significantly present in the extragalactic cosmic-ray flux, which seems to be quite improbable.

On the other hand, in the case of a heavy composition at the source, the UHECR flux is expected to drop dramatically above $3\,10^{20}$~eV, and would even be completely out of any foreseeable detection if the maximum energy at the sources could not provide secondary protons above $3\,10^{20}$~eV, i.e., if it were lower than $A\times 3\,10^{20}$~eV (even in this case, the flux would presumably be very low since the secondary protons would originate from nuclei accelerated at much higher energies). The implications of the two different classes of assumptions about the source composition thus result in totally opposite observational perspectives for the next generation of UHECR experiments. Measurements with larger statistics and a better understanding of the systematic uncertainties are needed to distinguish between the two scenarios.

\subsection{Low $E_{max}$ proton solutions}

A couple of arguments can however be used to claim that heavy source compositions are actually disfavored. First, hydrogen is the most abundant element in the universe and it is likely to be most abundant in the case of cosmic-rays accelerated to ultra-high energies (although  acceleration mechanisms are very poorly constrained). Second, the composition analysis at lower energy seem to favor a transition from heavy galactic to light extragalactic cosmic-rays which does not look compatible with the expectations of an extragalactic heavy source composition (see discussion of the pure iron sources in \cite{Denis2007a}). Let us note that the latter point is somehow alleviated but remains true if one assumes a strong evolution of source luminosity with redshift for which spectral indices of 2.0-2.1 would be required to fit the observed spectra. 

Both arguments can however be countered if one assumes that the composition at the source is actually proton dominated but that the proton maximum energy is lower than energy of the GZK feature leading to a heavy dominated composition at the highest energies (this kind of scenario is proposed  for instance in \cite{Inoue2007}). 

\begin{figure*}[ht]\centering\hfill~\hfill\includegraphics[width=0.47\linewidth]{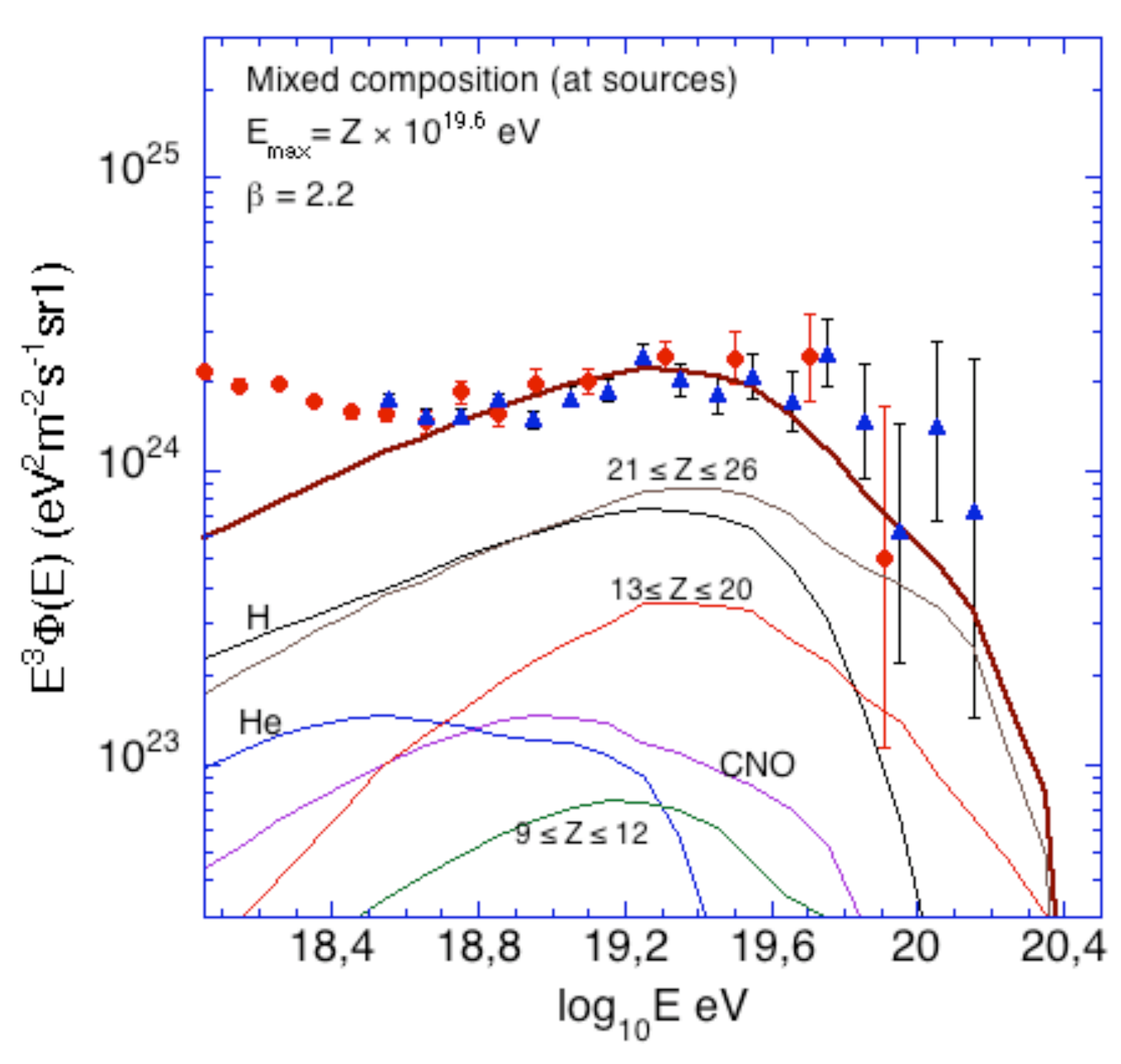}
\hfill\includegraphics[width=0.47\linewidth]{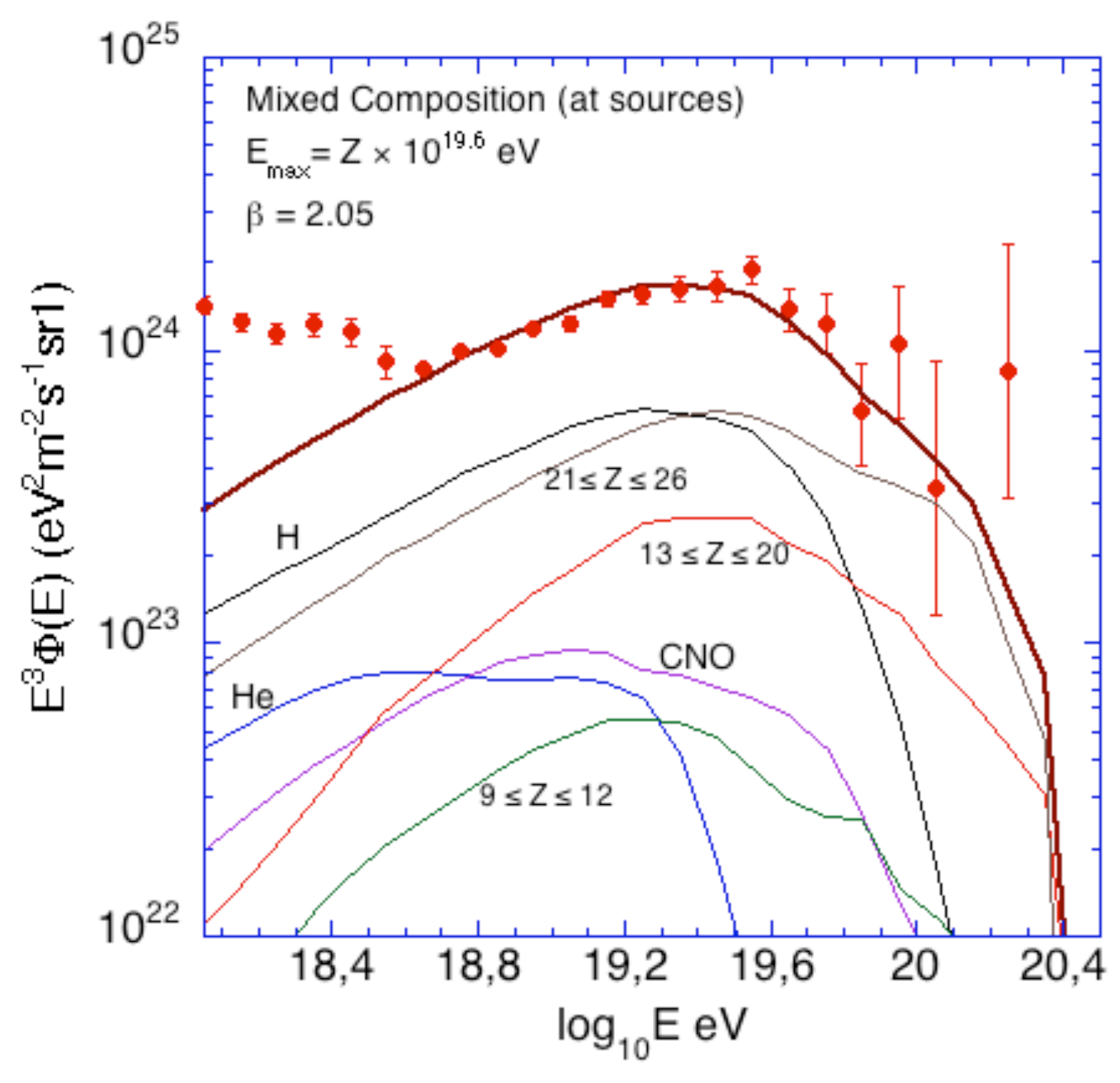}\hfill~\caption
{Propagated spectra obtained assuming a mixed source composition and a low proton maximum energy at the accelaration compared to HiRes (left) and Auger (right) spectra.}\label{LowEmaxSpectra}\end{figure*}

As we discussed in \cite{Denis2007a}, low $E_{max}$ proton solutions do not work very well with our usual mixed composition hypothesis. Indeed, at the sources the composition is assumed to be dominated by protons, with a large abundance of He nuclei and CNO and a lower abundance of heavy nuclei. In this case, the early cut-off of the He and CNO components  (which are not masked by secondary protons for low $E_{max}$ hypotheses) that closely follow the cut-off of protons (due to the maximum energy at the source) result in a sharp cut-off that is incompatible with data if protons are not acceleration above the GZK effect energy threshold. Some tuning of the composition is then necessary for this type of scenario to be compatible with  the data. However, acceptable fits of the data can be obtained by assuming that the heavy nuclei are more abundant than He and CNO at the source. This is illustrated in Fig.~\ref{LowEmaxSpectra} where expected spectra are displayed and compared with data assuming a mixed composition, $E_{max}=Z\times4\,10^{19}$ eV and $\sim30\%$ of Fe nuclei at the sources. One can see that the agreement with data is reasonable (especially with Auger spectrum) and that the composition, proton dominated at low energy, becomes gradually heavier and very dominated by iron above $5\,10^{19}$ eV. The implication for the flux above $3\,10^{20}$ eV are basically the same as in the pure iron source composition seen before (the main difference being that in the $E_{\max}$ case, one does not expect any secondary protons after the heavy component final drop). Cosmogenic neutrino fluxes at high energy (above $10^{17}$ eV) would as well be hopelessly low in this case. Indeed, pion production from the interactions of either  nucleons or nuclei with CMB photons would be highly suppressed for the low Lorentz factors implied by the low values of $E_{\max}$ \cite{Denis2006}.

\subsection{$X_{max}$ measurements}
As we pointed out in \cite{Denis2007a,Denis2007b,Denis2008} $ X_{\max}$ measurements will be as well a precious observable to distinguish between models. Indeed, In the case of proton dominated compositions at the highest energy the spectrum would be of no help to distinguish between compatible source compositions which could be either pure proton, proton dominated, He dominated, or intermediate mas nuclei dominated source compositions. As we showed in previous studies, the shape of the $X_{\max}$ evolution clearly removes the degeneracy between pure proton and proton dominated mixed composition models which would  basically give identical spectra at the highest energies. In the case of He or CNO dominated compositions, one also expects to obtain a distinct shape as most protons in the composition would be secondary nucleons. As seen in Fig.~\ref{Secondary} the composition would get continuously lighter and would not exhibit any \emph{delay of the lightening} feature  \cite{Denis2007a, Denis2007b} characteristic of proton dominated mixed compositions. Let us note finally that the low $E_{max}$  solutions we studied in the previous section would have a flatter evolution above the ankle than the usual mixed composition case as the composition above $10^{19}$ eV would get heavier instead of lighter.

As mentioned in previous studies  \cite{Denis2007a, Denis2007b, Denis2008}, Fly's Eye \cite{Bird+93} and HiRes \cite{HiRescomp} data show a good agreement proton dominated mixed composition predictions. In the case of Auger data, though the agreement is quite good with previous experiments, the evolution of $X_{\max}$ appears to be a bit flatter above the ankle. It is of course too early to conclude wether or not this suggests a composition getting heavier at the highest energies and further statistics, as well as cross checks of the predictions of hadronic models \cite{K07, KG07, Ralph07} will be help understand the energy evolution of $X_{max}$ at the highest energies.

\subsection{Conclusion}

In this paper, we have shown that current experimental spectra are compatible either with the standard GZK cut-off  of  proton dominated UHECRs  or with the "nuclear GZK cut-off" coming from the photodisintegration of  heavy dominated UHECRs.
In both cases, light and intermediate mass nuclei are not expected to play any significant role above $\sim5\,10^{19}$ eV due to their interaction with the photon backgrounds even if they were present or even dominant at the sources. Proton or heavy dominated compositions result in different shapes of the high energy cut-off of the ultra-high energy spectrum. Though current data do not allow to distinguish between them, it is likely that data  with higher statistics and lower systematics, as expected from the Pierre Auger observatory \cite{PAO}, will provide better constraints in the near feature. In the meantime, direct composition studies with higher statistics as well could provide important clues to solve this important issue.

Note that both classes of scenarios have different implications for the expected flux above $3\,10^{20}$ eV. In particular, if the maximum proton energy at the sources is large enough, a quasi pure proton composition is expected in this energy range, potentially extending well above $3\,10^{20}$ eV due to the contribution of nearby sources. Conversely, extremely low fluxes above $3\,10^{20}$ eV  would result from a heavy-dominated composition notably in the case of a low proton $E_{\max}$. Likewise, the secondary cosmogenic neutrino fluxes would be extremely low in this case. The discrimination between the two scenarios, with better measurements of the spectral shape and the energy scale of the high energy cut-off, as well as the corresponding $X_{\max}$ evolution, is then extremely important to constrain the acceleration models and understand the origin of UHE cosmic-rays.

\end{document}